\documentclass[useAMS,usenatbib,fleqn]{mn2e}
\usepackage{times}
\usepackage{amsmath}
\usepackage{xfrac}
\usepackage{amssymb}
\usepackage{float}
\usepackage{stfloats} 	
\usepackage{subfig}
\usepackage{balance}
\usepackage{graphicx}
\graphicspath{{./Figures/}}

\title[Bias of faint radio sources]{Evolution in the bias of faint radio sources to $\bf{z\sim2.2}$}
\author[S.~N.~Lindsay, M.~J.~Jarvis and K.~McAlpine]
{\parbox{\textwidth}{S.~N.~Lindsay,$^1$\thanks{E-mail: s.lindsay2@herts.ac.uk} 
M.~J.~Jarvis,$^{2,3}$ and 
K.~McAlpine$^3$
} 
\vspace{0.4cm}\\ 
\parbox{\textwidth}{
$^1$Centre for Astrophysics Research, Science \& Technology Research Institute, University of Hertfordshire, AL10 9AB, UK \\
$^2$Oxford Astrophysics, Department of Physics, Keble Road, Oxford, OX1 3RH, UK\\
$^3$Physics Department, University of the Western Cape, Bellville 7535, South Africa \\
}}
\begin{document}

\date{\today}

\pagerange{\pageref{firstpage}--\pageref{lastpage}} \pubyear{2014}

\maketitle

\label{firstpage}

\begin{abstract}
Quantifying how the baryonic matter traces the underlying dark matter distribution is key to both understanding galaxy formation and our ability to constrain the cosmological model.
Using the cross-correlation function of radio and near-infrared galaxies, we present a large-scale clustering analysis of radio galaxies to $z\sim2.2$. We measure the angular auto-correlation function of $K_\textrm{s}<23.5$ galaxies in the VIDEO-XMM3 field with photometric redshifts out to $z=4$ using VIDEO and CFHTLS photometry in the near-infrared and optical. We then use the cross-correlation function of these sources with 766 radio sources at $S_{1.4} > 90$ $\mu$Jy to infer linear bias of radio galaxies in four redshift bins. We find that the bias evolves from $b = 0.57\pm0.06$ at $z\sim0.3$ to $8.55\pm3.11$ at $z\sim2.2$. Furthermore, we separate the radio sources into subsamples to determine how the bias is dependent on the radio luminosity, and find a bias which is significantly higher than predicted by the simulations of Wilman et al., and consistent with the lower luminosity but more abundant FR-\textsc{I} population having a similar bias to the highly luminous but rare FR-\textsc{II}s. Our results are suggestive of a higher mass, particularly for FR-\textsc{I} sources than assumed in simulations, especially towards higher redshift.

\end{abstract}


\section{Introduction}

The observed distributions of galaxies and clusters today are far removed from the homogeneous picture we have of the early Universe using the cosmic microwave background (CMB; e.g. \citealt{komatsu11}), and we require large numbers of them to piece together a statistical understanding of clustering on cosmological scales. Cosmological applications of large-scale clustering measurements require information about the gravitating mass distribution in the Universe, which in a $\Lambda$CDM cosmology is strongly tied to the dark matter distribution. Direct observations tell us only about the baryonic matter, from which we must infer the dark matter distribution. Various tools exist for measuring the clustering signal of observed sources, such as nearest neighbour measures (e.g. \citealt{bahcall83}), counts-in-cells (e.g. \citealt{magliocchetti99,blake02b,yang11}), correlation functions (e.g. \citealt{groth77,bahcall83,blake02a,blake02b,croom05}) and power spectra (e.g. \citealt{cole05,percival07,komatsu11}).

Due to its relative simplicity to calculate, and relation to its Fourier transform (the power spectrum), the two-point spatial correlation function has become a standard in quantifying cosmological structure. A means by which we can quantify the extent to which the observable and dark matter are tied using the correlation function is through the linear bias parameter $b(z)$, the ratio of the galaxy correlation function to that of the dark matter. The bias quantifies the difference in the clustering of the dark matter haloes acting solely under gravity and of galaxies inhabiting those haloes with other effects making their structure more or less diffuse. This has a heavy dependence on the galaxy masses and the epoch under consideration (e.g. \citealt{seljak04}).

Extragalactic radio sources make useful probes of large-scale structure, being readily detectable up to high redshifts ($z\sim6$). Being unaffected by dust extinction, radio surveys are able to provide unbiased samples of larger volumes than would be available to an optical survey. Unfortunately many radio sources, particularly at higher redshifts, have very faint optical counterparts, which combines with the often extended nature of radio emission from active galactic nuclei (AGN) to make it difficult to optically identify and obtain redshifts for these individual sources \citep[see e.g.][]{mcalpine12}.

Without knowing the distance to a given radio source, clustering analyses are confined to two dimensions with the angular correlation function measuring any excess of source pairs as a function of their angular separation. The broad redshift distribution typical of radio surveys can make detection of this clustering difficult as the majority of close pairs of sources are widely separated in the line of sight direction, diluting any genuine clustering signal. Strong detections of this clustering signal over a range of angular separations were made possible with the advent of large area radio surveys observing to depths of a few mJy: e.g. Faint Images of the Radio Sky at Twenty-cm (FIRST; \citealt{becker95}), the Westerbork Northern Sky Survey (WENSS; \citealt{rengelink97}), the NRAO VLA Sky Survey (NVSS; \citealt{condon98}) and the Sydney University Molonglo Sky Survey (SUMSS; \citealt{bock99}). A positive correlation function is measured with high significance using these surveys extending to separations of several degrees (see work by \citet{cress96} and \citet{magliocchetti98,magliocchetti99} with FIRST, \citet{blake02a,blake02b} and \citet{overzier03} with NVSS, \citet{rengelink98} with WENSS, and \citet{blake04} with SUMSS).

Angular clustering studies (such as those above) assume a power law form for the spatial correlation function (also a common assumption for direct spatial clustering measurements; e.g. \citealt{magliocchetti04,brand05}), which is also preserved in angular projection \citep{limber53}. Inferring the spatial clustering properties of a galaxy sample from this projected form requires a knowledge of the redshift distribution of the sample, itself subject to uncertainty in addition to that resulting from the diluted signal arising from a sample with a broad redshift distribution. Photometric redshift surveys, while lacking the precision obtained from galaxy spectra necessary for a full 3D clustering analysis, provide a more accurate redshift distribution for a given sample of radio sources than assumed models or luminosity functions. Furthermore, they allow a sample to be divided into redshift bins, each with a well known distribution (given large enough numbers to account for small photometric errors). \citet{lindsay14} use a combination of spectroscopic and photometric redshifts to investigate the clustering of FIRST radio sources to $z\sim 1.5$ and provide some comparison between spatial and angular correlation function results (with and without the precision of spectroscopic redshifts, respectively). However, redshift measurements are lacking at high redshift where the clustering is stronger but poorly constrained over such large areas.

While sky coverage alone, with surveys such as NVSS, provides the statistical power required to measure the strength of the clustering of radio sources over large scales to depths of a few mJy, the depths of similarly large-area optical surveys, spectroscopic or otherwise, do not allow for optical identification of the radio sources with any significant completeness. However, small-area surveys have been carried out at the $\mu$Jy level, which also have potential uses for cosmology and large-scale structure measurements. At the $\mu$Jy level, the radio population becomes less dominated by FR-\textsc{I} and FR-\textsc{II} type active galactic nuclei (AGN), and we observe a greater fraction of star-forming galaxies. However, the lower flux-density limit also extends the range at which we can detect AGN, reaching beyond $z\sim1$ where the bias of radio sources is poorly understood. It is important to measure the bias of the radio sources to these high redshifts, and to know how it evolves, in order to inform cosmological experiments dependent on disentangling the observed galaxy clustering from other effects, such as cosmic magnification (e.g. \citealt{scranton05, wang11}) and the integrated Sachs-Wolfe effect (ISW; \citealt{mcewen07, giannantonio08, raccanelli08}). In particular the large volume of the Universe that will be opened up by the SKA and its precursors may provide important information on the very largest scales \citep[e.g.][]{raccanelli12,camera12}.

The aim of this paper is to investigate the bias of a sample of faint radio sources, extending to $z>2$ where observational measurements are lacking. We use 1.4 GHz radio data from the VLA-VIRMOS Deep Field \citep{bondi03} covering an area of 1 square degree to $S_{1.4} > 90$ $\mu$Jy, overlapping with optical photometry from the Canada-France-Hawaii Telescope Legacy Survey (CFHTLS) Deep-1 field (D1) and near-infrared photometry from the VISTA Deep Extragalactic Observations (VIDEO; \citealt{jarvis13}) survey with a depth of $K_\textrm{s} < 23.5$.  We overcome the signal-to-noise issue of having a far smaller sample than a wider, shallower NVSS-like survey by inferring the properties of the radio sources by the combined use of the angular correlation function of $K_\textrm{s}$-selected VIDEO	 sources (with sufficiently large numbers to keep uncertainties small) and the angular cross-correlation of these VIDEO sources with the radio sample \citep[see e.g.][for similar use of this technique]{guo11,hartley13}.
With reliable photometric redshifts out to $z\sim4$ for all of the galaxies used, we have a good knowledge of the redshift distributions of our samples as well as estimates of their radio luminosity. Even when coarsely binning by redshift, this gives us valuable constraints on the bias of these radio sources in bins up to a median redshift of $z=2.15$, and an insight into the clustering specifically of typical radio AGN at high redshift.

This paper is organized as follows: Section \ref{data} describes the multi-wavelength surveys from which we construct our galaxy samples. Section \ref{methods} details the correlation function methods used to calculate the galaxy bias and Sections \ref{results} and \ref{discussion}, respectively, show our results and present our discussion of them. The results are summarized in Section \ref{conclusions}.

The cosmological model used throughout this paper is the flat, $\Lambda$CDM concordance cosmology where $\Omega_m = 0.3$, $\Omega_\Lambda = 0.7$ and $\sigma_8 = 0.8$. All distances are kept in units of $h^{-1}$Mpc where $H_0 = 100 h$ km s$^{-1}$Mpc$^{-1}$ and $h$ is not explicitly assumed.

\section{Data}\label{data}

\subsection{Near-Infrared observations}\label{VIDEO}
	
Our near-infrared galaxy catalogue comes from the Visible and Infrared Survey Telescope for Astronomy (VISTA) facility in Chile. This is a 4.1m wide-field survey telescope with a 1.65-degree field of view and a 67-megapixel near-infrared camera. The VISTA Deep Extragalactic Observations (VIDEO; \citealt{jarvis13}) survey covers $\sim$12 square degrees over three fields in five near-infrared bands, tracing the evolution of galaxies and large-scale structure from the present out to $z=4$, and higher for AGN and the most massive galaxies. The survey photometry reaches $5\sigma$ AB-magnitude depths of 25.7, 24.5, 24.4, 24.1 and 23.8 (in 2 arcsec apertures) in \textit{Z, Y, J, H} and \textit{K$_s$} bands, respectively. 
	
	The VIDEO-XMM3 tile detailed in \citet{jarvis13} overlaps with the Canada-France-Hawaii Telescope Legacy Survey (CFHTLS; \citealt{ilbert06}) Deep-1 field (D1) optical data (\textit{u*, g', r', i', z'} bands) over a 1 x 1 degree area. The combination of these optical and near-infrared data allow for improved photometric redshifts with 3.1 per cent catastrophic outliers at $z<1$, and further improvements at $1<z<4$ expected due to VIDEO's sensitivity to the Balmer and 4000\AA \ breaks at these redshifts (see \citealt{jarvis13} for further details).
	
\subsection{Radio observations}\label{radio}
	
	The radio data used in this analysis come from Very Large Array (VLA) observations at 1.4 GHz by \citet{bondi03} in the VLA-VIRMOS Deep Field, corresponding to the same 1 deg$^2$ area covered by both VIDEO and CFHTLS-D1. This survey used the VLA in B-configuration, giving approximately 6 arcsec resolution, producing a final mosaic image of nearly uniform noise at $\simeq$ 17 $\mu$Jy.
	
	\citet{mcalpine12} used a likelihood ratio (LR) method \citep{sutherland92} to identify infrared counterparts to radio sources in the VLA-VIRMOS Deep Field. Of the 1054 5$\sigma$ detections, 915 were found to have reliable (Rel $>0.8$) counterparts in VIDEO. These are a combination of radio-loud AGN, quasars and star-forming galaxies \citep[see e.g.][]{mcalpine13}.
	
\subsection{Final Samples}
	
	Due to particularly bright sources in the VIDEO images adding to the noise in surrounding areas and obscuring faint nearby sources, we apply a mask to our parent near-infrared catalogue. A circular area of 0.01\degr\ in radius is excised, centred on any sources brighter than 13.5 mag. 
	
\begin{figure}
  \centering
 \includegraphics[width=0.48\textwidth]{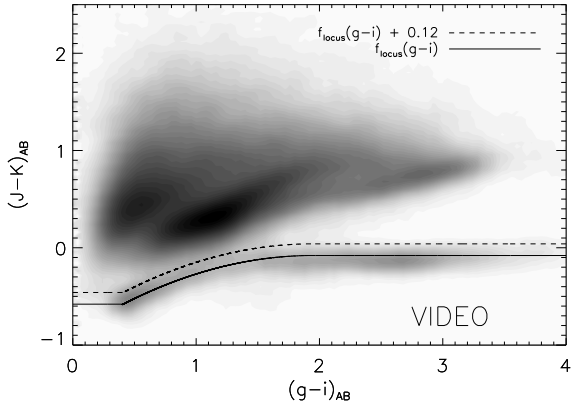}
   \caption{\textit{J--K} vs. \textit{g--i} colours of the VIDEO sources ($K_\textrm{s}<23.5$), with the stellar locus (\emph{solid line}) and imposed galaxy cut-off (\emph{dashed line}) as used by \citet{mcalpine12}.}
  	\label{stargalsep}
\end{figure}

	Any remaining stellar contaminants are removed from the VIDEO catalogue by the same means used in \citet{mcalpine12} following the method of \citet{baldry10}.  We use the \textit{J--K} and \textit{g--i} colours to define a stellar locus, shown in Figure \ref{stargalsep}. All sources more than 0.12 mag redward in \textit{J--K} of this stellar locus are considered galaxies and remain in our VIDEO sample. 
	
	Finally, we impose a magnitude limit of $K_\textrm{s} < 23.5$ to the near-infrared catalogue, corresponding roughly to the 22.6 (Petrosian magnitude) cut-off applied in the radio cross-matching, and a flux density limit of $S_{1.4} >$ 90 $\mu$Jy to the radio sources. While the nominal detection limit of \citet{bondi03} is 80 $\mu$Jy, with some 41 sources detected at lower flux densities down to $\sim$ 60 $\mu$Jy, there is still appreciable incompleteness at this limit which merits the slightly more conservative cut used in our analysis. This results in a final sample of 766 objects in our radio sample, and 95,826 in the corresponding infrared sample.

\section{Methods}\label{methods}
\subsection{Angular Correlation Function}

The angular two-point correlation function, $w(\theta)$ is defined as the excess probability of finding a galaxy at an angular distance $\theta$ from another galaxy, as compared with a Poissonian (unclustered) distribution \citep{peebles80}:
\begin{equation}
	\delta P = \sigma[1 + w(\theta)]\delta\Omega ,
\end{equation}
where $\delta P$ is the probability, $\sigma$ is the mean surface density and $\delta\Omega$ is the surface area element. This quantifies the degree of clustering apparent at a given angular scale, with $w(\theta)$ generally decreasing monotonically with increasing $\theta$ as gravitational interactions become weaker at large separations.
	
	Estimators of $w(\theta)$ use the measured quantities of $DD(\theta)$ and $RR(\theta)$ which represent the number of galaxy pairs separated by $\theta$ in the real data and a corresponding random catalogue, respectively. The cross-pair separations $DR(\theta)$ are used in slightly more sophisticated estimators, reducing the variance, as in our chosen estimator by \citet{landy93}:
	\begin{equation}
	w(\theta) = \frac{DD}{RR} - 2\frac{DR}{RR} + 1 ,
	\label{LS}
\end{equation}	
where each quantity is normalized such that its sum over all values of $\theta$ is unity.

By averaging over several random data sets and using $\overline{DR}$ and $\overline{RR}$, or by using a more densely populated random catalogue, we may assume the statistical error in the random sets to be negligible. The random catalogues themselves have been autocorrelated finding no significant deviation from zero even at extremes of angular separation where the counts are low. The variance of the correlation function, therefore, is often given by the Poisson error due to the DD counts alone
\begin{equation}
	\Delta w = \frac{1 + w(\theta)}{\sqrt{DD}}.
\end{equation}
	
However, the errors in the correlation function depend on the $DD$ counts beyond simple Poisson variance; adjacent bins are correlated, with each object contributing to counts across a range of separation bins. The errors are therefore calculated more rigorously using a bootstrap resampling technique \citep{ling86} whereby several data catalogues are constructed by randomly sampling (with replacement) the original set of objects. As such, in any given set, some sources are counted twice or more and some not at all. The resulting binned $DD$ counts should give a mean approximately equal to the original data but allow us to calculate a variance for each bin, and therefore $w(\theta)$ values. \citet{lindsay14} found errors in $w(\theta)$ for a subset of FIRST galaxies in the GAMA survey using a bootstrap resampling method. Their Poisson error estimates were consistently a factor of 1.5--2 smaller than the bootstrap error up to $\sim$ 0.02\degr, above which the ratio increased rapidly with $\theta$. 

The restricted survey area from which we can measure $w(\theta)$ results in a negative offset in the observed correlation function, known as the \emph{integral constraint}. Expressed mathematically, the relation between observed correlation function $w_\textrm{obs} (\theta)$ and the genuine function $w(\theta)$ is
\begin{equation}
w_\textrm{obs} (\theta) = w(\theta) - \sigma^2 ,
\end{equation}
where $\sigma^2$ represents the integral constraint \citep{groth77} which can be approximated, following \citet{roche99}, by
\begin{equation}
\sigma^2 = \frac{\sum RR(\theta) w(\theta)}{\sum RR(\theta)} .
\end{equation}

\begin{figure}
  \centering
    \includegraphics[width=0.48\textwidth]{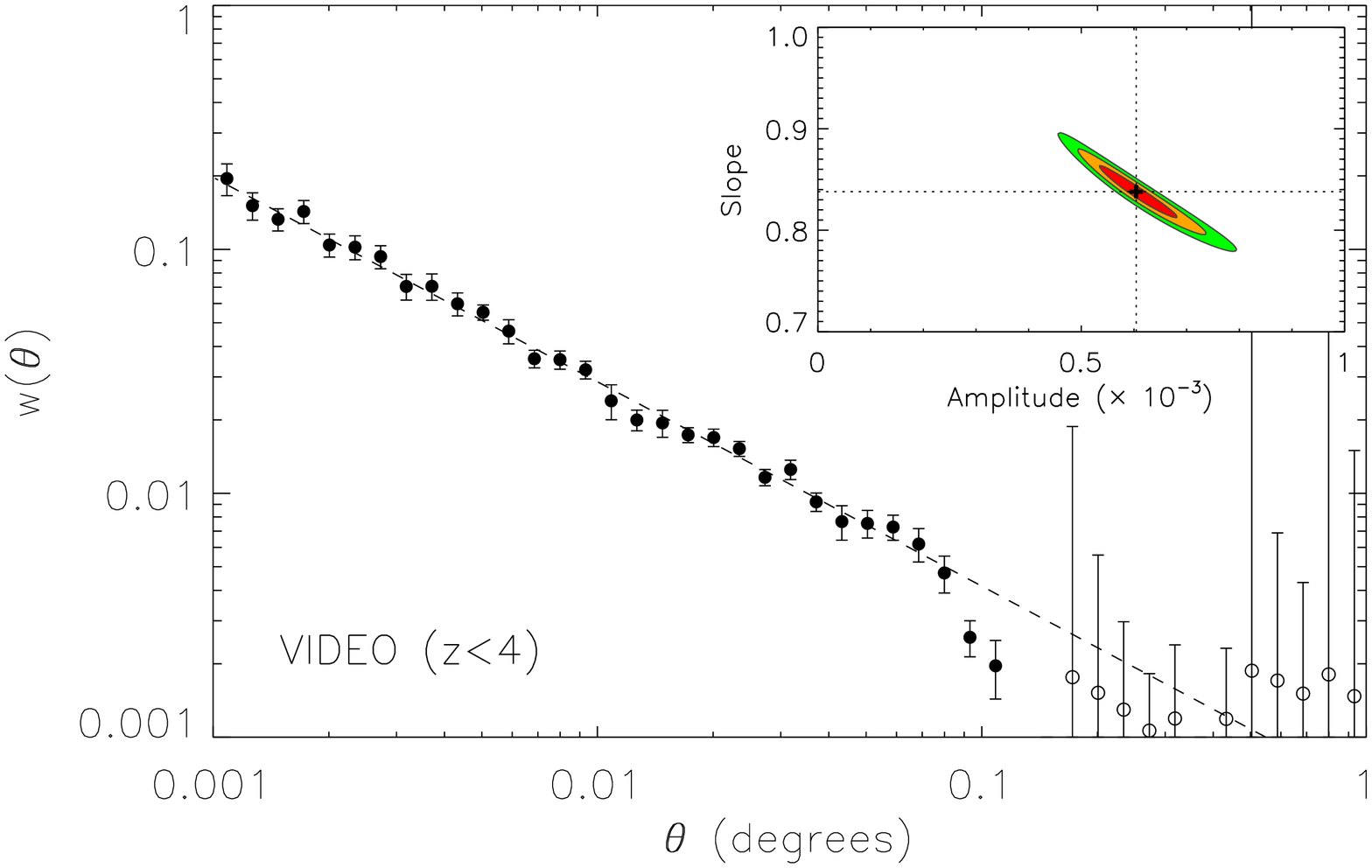}
   \caption{The angular correlation function for $K_\textrm{s}<23.5$ near-infrared galaxies (with bootstrap resampling errors). The dashed line shows the best fit power-law (for $\theta < 0.8$) and the inset shows parameter fits at 68, 90 and 95 per cent confidence levels. The open points show the absolute values where $w(\theta) < 0$.}
 \label{wIR}
\end{figure}

Traditionally, $w(\theta)$ is fitted with a power law (e.g. Peebles, 1980) with slope $\sim 0.8$, commonly found for the clustering of objects selected in a variety of ways \citep[e.g.][]{bahcall83}.
In Figure~\ref{wIR} we show the angular correlation function of our $K_\textrm{s}$-selected VIDEO galaxies in order to understand their inherent large-scale clustering properties as a stepping stone towards investigating the relative clustering between these galaxies and the subset of radio sources in the same VIDEO field.

\subsection{Cross-correlation Function}

Closely related to the two-point correlation function is the two-point cross-correlation function, which compares two different data sets $D_1$ and $D_2$. The same procedure is followed as for the auto-correlation of one data set, but we modify Eq. \ref{LS} as follows:
\begin{equation}
	w_\textrm{cross}(\theta) = \frac{D_{1}D_{2} - D_{1}R - D_{2}R}{RR} + 1,
\end{equation}
where $D_{1}D_{2}$ and $D_{i}R$ are analogous to $DD$ and $DR$ above, and are similarly normalized.

\begin{figure}
  \centering
    \includegraphics[width=0.48\textwidth]{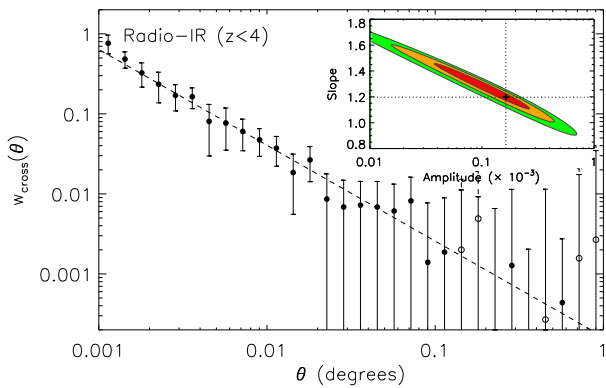}
   \caption{The angular cross-correlation function of the radio sources to the VIDEO infrared sources (with bootstrap resampling errors). The dashed line shows the best fit power-law and the inset shows $\chi^2$ parameter fits at 68, 90 and 95 per cent confidence levels.  The open points show the absolute values where $w_\textrm{cross} (\theta) < 0$.}
  	\label{wcross}
\end{figure}

The cross-correlation function is also fitted with a power law in the same manner as the auto-correlation function and describes the relative cross-clustering of two populations with one another. Here, we use the cross-correlation function of VIDEO $K_\textrm{s}$-selected galaxies with radio sources in order to provide a higher signal-to-noise measurement of the clustering than would be possible with the radio auto-correlation alone. Figure~\ref{wcross} shows the angular cross-correlation function for the radio sources with the near-infrared selected sources, which in the following sections we use to infer the clustering properties of the radio galaxies themselves.

\subsection{Spatial clustering and Limber inversion}

	The 3-dimensional analogue of $w(\theta)$ is the spatial two-point correlation function $\xi(r )$, which measures the excess probability, due to clustering, of finding a pair of objects separated by $r\rightarrow r + \delta r$ as compared with a Poissonian (unclustered) distribution, similarly defined:
\begin{equation}
	\delta P = n[1 + \xi(r )]\delta V ,
\end{equation}
where $n$ is the mean number density of objects and $\delta V$ a volume element.

The two-point correlation function is usually fitted with a single power law over a significant range of separations as follows:
	
\begin{equation}
	\xi(r ) = \left(\frac{r_0}{r}\right)^{\gamma} \label{eq:xir} ,
\end{equation} 
with $r_0$ called the correlation length. This is not a physical length but represents the separation length at which $\xi(r ) = 1$. As the function increases toward smaller separations, this is approximately equivalent to the statement that $r_0$ is the length below which $DD(r ) > 2RR(r )$. For any reasonable slope, therefore, a larger $r_0$ implies a more strongly clustered distribution. 

\begin{figure*}
  \centering
    \includegraphics[width=0.495\textwidth]{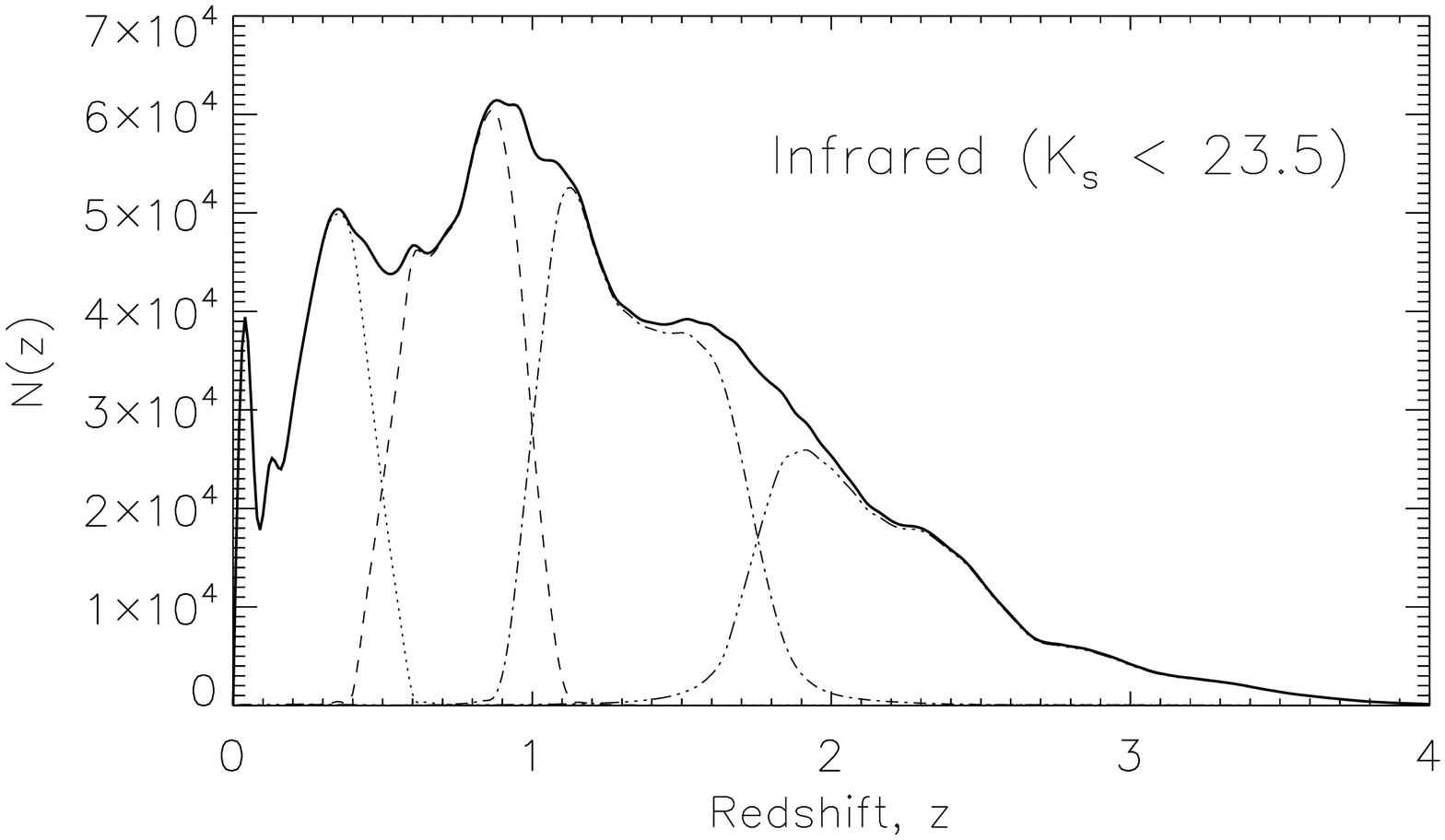}
    \includegraphics[width=0.495\textwidth]{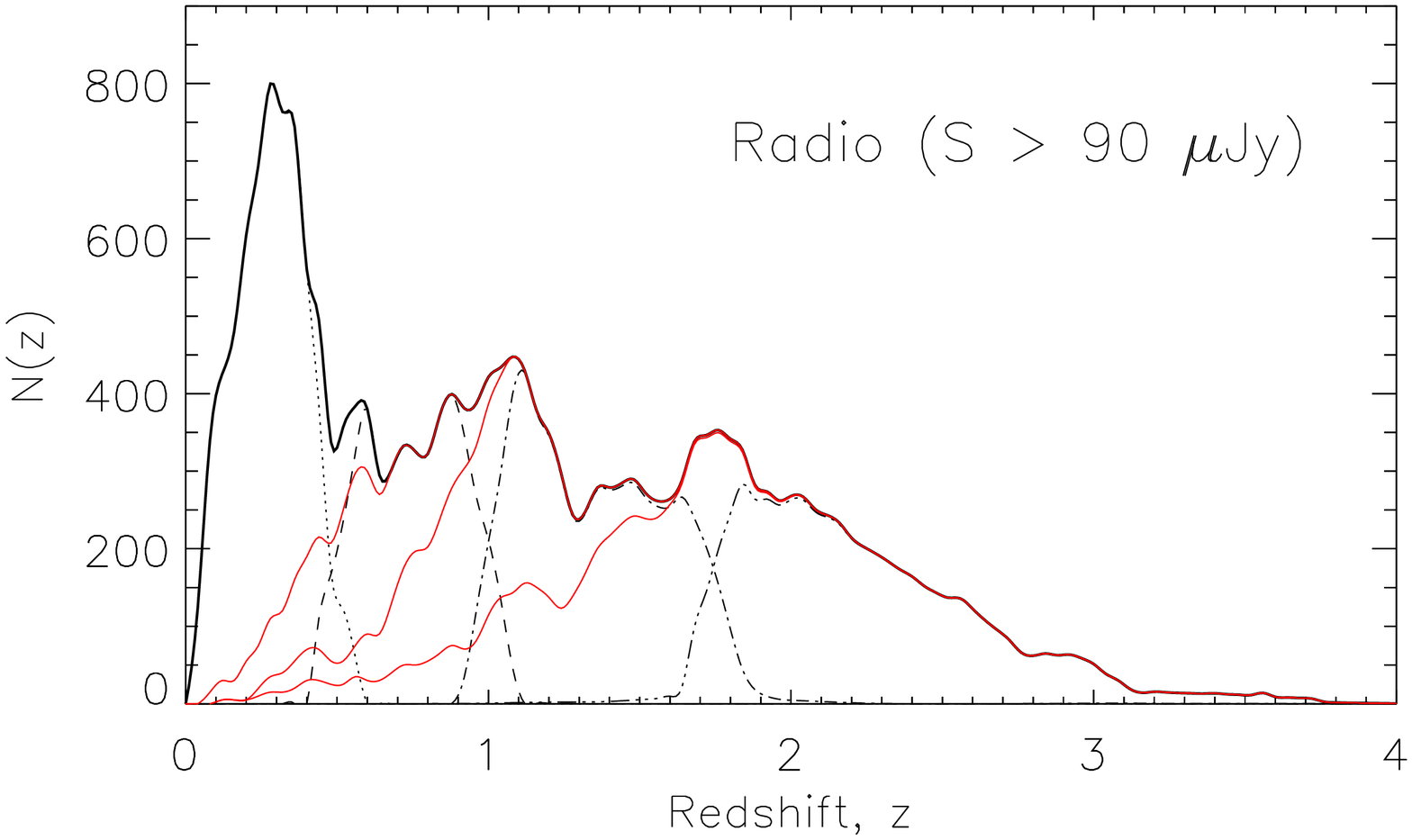}
   \caption{Redshift distributions of infrared-selected sources (\textit{left panel}) and radio-selected sources (\textit{right panel}). The four broken lines show the distributions of sources in our four redshift bins. The three red lines on the right panel show the distributions of radio sources with $L>10^{23}$, $10^{23.5}$ and $10^{24}$ WHz$^{-1}$ in increasing order of median redshift.}
  	\label{zbins}
\end{figure*}

%
%

If the redshift distribution of a set of objects is known, one may deproject the angular correlation function into the spatial correlation function. This is the purpose of the cosmological Limber equation \citep{limber53,peebles80} for estimating the correlation length, $r_0$. Determining $\xi(r)$ directly is difficult, as a complete set of individual redshifts is rarely available for a given survey, thus requiring the redshift distribution to be estimated in order to deproject $w(\theta)$. Using the photometric redshifts available for the galaxies in the VIDEO survey, however, means we may apply a redshift distribution directly from the data. The photometric redshift distributions for both the $K_{s}$-selected galaxies and the radio galaxies are calculated by adding the normalized probability distributions of the individual photometric redshifts of each object. This ensures that we fully incorporate the uncertainties associated with the photometric redshifts when determining the clustering of the various populations. These redshifts distributions are shown in Figure \ref{zbins}.

An epoch-dependent form of the spatial correlation function is assumed \citep[see e.g.][and references therein]{dezotti90,overzier03}:
\begin{equation}
	\xi(r,z) = \left(\frac{r_0}{r}\right)^{\gamma} \times (1+z)^{\gamma-(3+\epsilon)} ,
\end{equation}
where $\epsilon$ parameterises the clustering model being assumed. Following similar work in \citet{lindsay14}, we assume the comoving model with $\epsilon=\gamma-3$.

The spatial correlation function slope, $\gamma$, is the same as that used in the power-law fit to the angular correlation function (where the magnitude of the slope is $\gamma - 1$), so we measure this parameter through the $w(\theta)$ function. The amplitude $A$ of $w(\theta)$ has been expressed as a function of $r_0$ (in comoving coordinates) in the literature \citep{overzier03,kovac07,kim11,elyiv12} as follows:
\begin{equation}
A = r_0^\gamma H_\gamma \left(\frac{H_0}{c}\right) \frac{\int_0^\infty N^2(z) (1+z)^{\gamma - (3+\epsilon)} \chi^{1-\gamma}(z) E(z) \mathrm{d}z}{\left[ \int_0^\infty N(z) \, \mathrm{d}z \right]^2},
\label{eq:Limber}
\end{equation}
where  $H_\gamma =  \Gamma(\frac{1}{2}) \Gamma(\frac{\gamma-1}{2})/\Gamma(\frac{\gamma}{2})$, $N(z)$ is the redshift distribution and $\chi(z)$ is the comoving line-of-sight distance to an object at a redshift $z$:
\begin{equation}
\chi (z) = \frac{c}{H_0} \int^{z}_{0} \frac{\mathrm{d}z^{\prime}}{E(z^{\prime})}.\label{eq:comoving}
\end{equation}
Here, $H_0$ is the Hubble constant and $E(z)$ is the function used to describe the cosmological expansion history:
\begin{equation}
E(z) = \left[\Omega_{m,0} (1+z)^3 + \Omega_{k,0} (1+z)^2 + \Omega_{\Lambda,0}\right]^\frac{1}{2}. \label{eq:E(z)}
\end{equation} 
Equation \ref{eq:Limber} may simply be inverted to give the comoving correlation length, $r_0$ as a function of the redshift distribution, the correlation function slope and the angular clustering amplitude.

%
%
\subsection[]{Mass Bias}

	The differences in clustering of different classes of extragalactic objects and the background matter distribution motivates the use of some bias parameter, as introduced by \citet{kaiser84} and \citet{bardeen86}: 
	\begin{equation}
		b^2(z) = \frac{\xi_{\textrm{gal}}(r,z)}{\xi_{\textrm{DM}}(r,z)}, \label{eq:bias}
	\end{equation}

	where the numerator and denominator are the galaxy and dark matter correlation functions, respectively. 
	
	The bias parameter (as a function of redshift) may be defined as equation \ref{eq:bias} with $r=8 h^{-1}$Mpc. As equation \ref{eq:xir} shows, the numerator can be written
	\begin{equation}
	 \xi_{\textrm{gal}} (8,z) = \left[ \frac{r_0 (z)}{8} \right]^\gamma .
	 \end{equation}
	 The corresponding function for the denominator is given by \citet{peebles80} as
	 \begin{equation}
	  \xi_{\textrm{DM}} (8,z) = \sigma_8^2 (z) / J_2
	  \end{equation}
	  where $J_2 = 72/[(3-\gamma)(4-\gamma)(6-\gamma)2^\gamma]$ and the parameter $\sigma_8^2$ is the dark matter density variance in a comoving sphere of radius 8 $h^{-1}$Mpc. 
	  The combination of these equations gives the evolution of bias with redshift, given only the correlation length and slope:
	  \begin{equation}
	  	b(z) = \left[ \frac{r_0(z)}{8} \right]^{\gamma / 2} \frac{J_2^{1/2}}{\sigma_8 D(z) / D(0)}.
	\end{equation}
	In each case, the redshift used for presentation purposes is the median of the distribution of objects.
	
	Analogous to equation \ref{eq:bias} for the cross-correlation we have
	\begin{equation}
		b_{K\textrm{R}}^2(z) = \frac{\xi_{K\textrm{R}}(r,z)}{\xi_{\textrm{DM}}(r,z)} = b_K(z) b_\textrm{R}(z), \label{eq:crossbias}
	\end{equation}
	where $K$ and R subscripts denote $K_\textrm{s}$-selected galaxies, and radio galaxies respectively. Given that we are able to calculate the bias of the near-infrared sources, $b_K(z)$, from their auto-correlation function, the radio bias may be given by
	\begin{equation}
		b_\textrm{R}(z) = \frac{b_{K\textrm{R}}^2(z)}{b_K(z)}. \label{eq:radiobias}
	\end{equation}
	
	To account for the fact that we are measuring these quantities for discrete samples with different redshift distributions where $z$ for corresponding bins is subtly different between the NIR and radio sources, we multiply the right hand side of equation \ref{eq:radiobias} by $D(z_K)/D(z_\textrm{R})$, which is a relatively small correction. This gives a final quantity, $b_\textrm{R} (z_\textrm{R})$, which describes the bias of the radio sources at the redshift of those radio sources:
		\begin{equation}
		b_\textrm{R}(z_\textrm{R}) = \frac{b_{K\textrm{R}}^2(z_\textrm{R})}{b_K(z_K)} \frac{D(z_K)}{D(z_\textrm{R})}. \label{eq:radiobias2}
	\end{equation}

\section{Results} \label{results}

\begin{table*}
\begin{center}
\caption{Angular clustering parameters from the autocorrelation function, $w(\theta)$, and the inferred correlation length and bias for our various $K_\textrm{s}$-selected galaxy samples.}
 \begin{tabular}{c c c c c c r} \hline
 Redshift range   & IR sources & $z_\textrm{med}$	 & $A$ ($\times 10^{-3}$) & $\gamma$ 				 	& $r_0$ ($h^{-1}$Mpc) 		& $b_K(z)$ 				\\ \hline
   $z < 4$        & 95,826     & 1.09 			 & $0.61	^{+0.05}_{-0.05}$ & $1.84^{+0.02}_{-0.02}$ 	& $2.79^{+0.08}_{-0.09}$ 	& $1.12^{+0.03}_{-0.03}$ \\ \hline
   $z < 0.5$      & 17,603     & 0.33 			 & $2.55^{+0.32}_{-0.30}$ & $1.85^{+0.03}_{-0.03}$ 	& $2.08^{+0.07}_{-0.07}$ 	& $0.59^{+0.02}_{-0.01}$ \\ [1ex]
   $0.5 < z < 1$  & 25,461     & 0.79 			 & $2.06^{+0.18}_{-0.17}$ & $1.85^{+0.02}_{-0.02}$ 	& $2.59^{+0.08}_{-0.09}$ 	& $0.91^{+0.02}_{-0.03}$ \\ [1ex]
   $1 < z < 1.75$ & 31,507     & 1.33 			 & $1.45^{+0.12}_{-0.11}$ & $1.85^{+0.02}_{-0.02}$ 	& $2.73^{+0.10}_{-0.10}$ 	& $1.21^{+0.04}_{-0.04}$ \\ [1ex]
   $1.75 < z < 4$ & 21,255     & 2.16 			 & $0.69^{+0.12}_{-0.10}$ & $1.98^{+0.03}_{-0.03}$ 	& $3.80^{+0.18}_{-0.19}$ 	& $2.23^{+0.12}_{-0.12}$ \\ \hline
  \end{tabular}
  \label{IRpar}
\end{center}	
\end{table*}

	\begin{table*}
\begin{center}
\caption{Angular clustering parameters from the cross-correlation function, $w_\textrm{cross}(\theta)$, and the inferred correlation length and bias for radio-IR cross-clustering. The inferred radio bias $b_R (z)$ is shown in the final column.}
 \begin{tabular}{c c c c c c c c r}	 \hline
 Redshift range 	& Luminosity (WHz$^{-1}$) 	& Radio sources 	& $z_\textrm{med}$ 	& $A$ ($\times 10^{-3}$) & $\gamma$ 				& $r_0$ ($h^{-1}$Mpc) 	& $b_{K\textrm{R}}(z)$ 	& $b_\textrm{R}(z)$ 	\\ \hline
   $z < 4$ 		& All 						& 766 			& 1.02 				& $0.17^{+0.14}_{-0.08}$ & $2.19^{+0.11}_{-0.11}$	& $3.83^{+0.21}_{-0.24}$	& $1.57^{+0.12}_{-0.15}$	& $2.13 \pm 0.27$ 	\\ \hline
   $z < 0.5$ 	& All 						& 234 			& 0.29 				& $0.99^{+0.47}_{-0.33}$ & $2.15^{+0.08}_{-0.08}$	& $2.57^{+0.16}_{-0.18}$	& $0.70^{+0.05}_{-0.05}$	& $0.57 \pm 0.06$ 	\\ [1ex]
   $0.5 < z < 1$ & " 						& 139 			& 0.75 				& $1.28^{+0.74}_{-0.54}$ & $2.07^{+0.11}_{-0.10}$	& $3.94^{+0.41}_{-0.48}$	& $1.38^{+0.20}_{-0.19}$	& $1.80 \pm 0.36$ 	\\ [1ex]
   $1 < z < 1.75$ & " 						& 194 			& 1.35 				& $0.70^{+0.59}_{-0.41}$ & $2.09^{+0.16}_{-0.13}$	& $4.62^{+0.68}_{-0.75}$	& $2.11^{+0.48}_{-0.39}$	& $4.09 \pm 1.20$ 	\\ [1ex]
   $1.75 < z < 4$ & " 						& 199 			& 2.15 				& $0.35^{+0.29}_{-0.20}$ & $2.20^{+0.16}_{-0.13}$	& $5.61^{+0.83}_{-1.00}$	& $3.60^{+0.99}_{-0.84}$	& $8.55 \pm 3.11$ 	\\ \hline
   $z < 4$ 		&$\log(L)>23.0$ 				& 575 			& 1.37 				& $0.16^{+0.11}_{-0.07}$ & $2.24^{+0.10}_{-0.10}$	& $5.57^{+0.33}_{-0.36}$	& $2.76^{+0.34}_{-0.30}$	& $7.62 \pm 1.27$ 	\\ [1ex]	
   	" 			&$\log(L)>23.5$ 				& 499 			& 1.55 				& $0.11^{+0.09}_{-0.06}$ & $2.28^{+0.13}_{-0.12}$	& $5.57^{+0.50}_{-0.52}$	& $3.04^{+0.61}_{-0.46}$	& $9.91 \pm 2.48$ 	\\ [1ex]
	"			&$\log(L)>24.0$ 				& 372 			& 1.77 				& $0.11^{+0.12}_{-0.06}$ & $2.28^{+0.15}_{-0.14}$	& $5.30^{+0.47}_{-0.61}$	& $3.10^{+0.64}_{-0.54}$	& $11.14 \pm 3.01$	\\  [1ex] 
   \hline
   
  \end{tabular}
  \label{crosspar}
\end{center}	
\end{table*}

\begin{figure}
  \centering
    \includegraphics[width=0.48\textwidth]{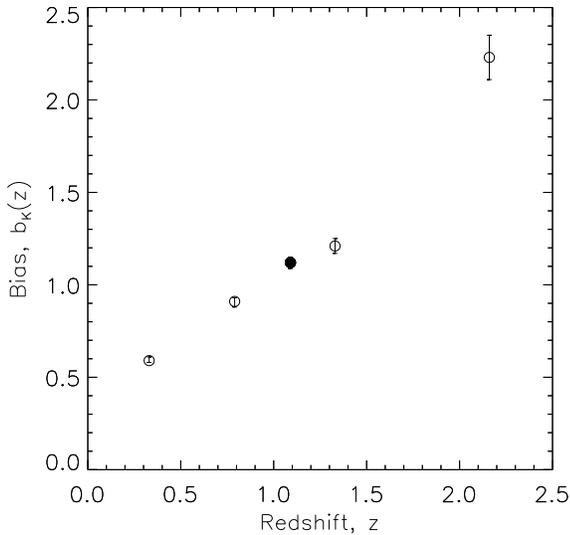}
   \caption{Linear bias of VIDEO $K_\textrm{s}$-selected sources as a function of median redshift. Open circles correspond to the four independent redshift bins used while the filled circle is the bias for the full sample of 95,826 sources with $K_\textrm{s} < 23.5$.}
  	\label{biasIR}
\end{figure}

\begin{figure*}
  \centering
    \includegraphics[width=0.9\textwidth]{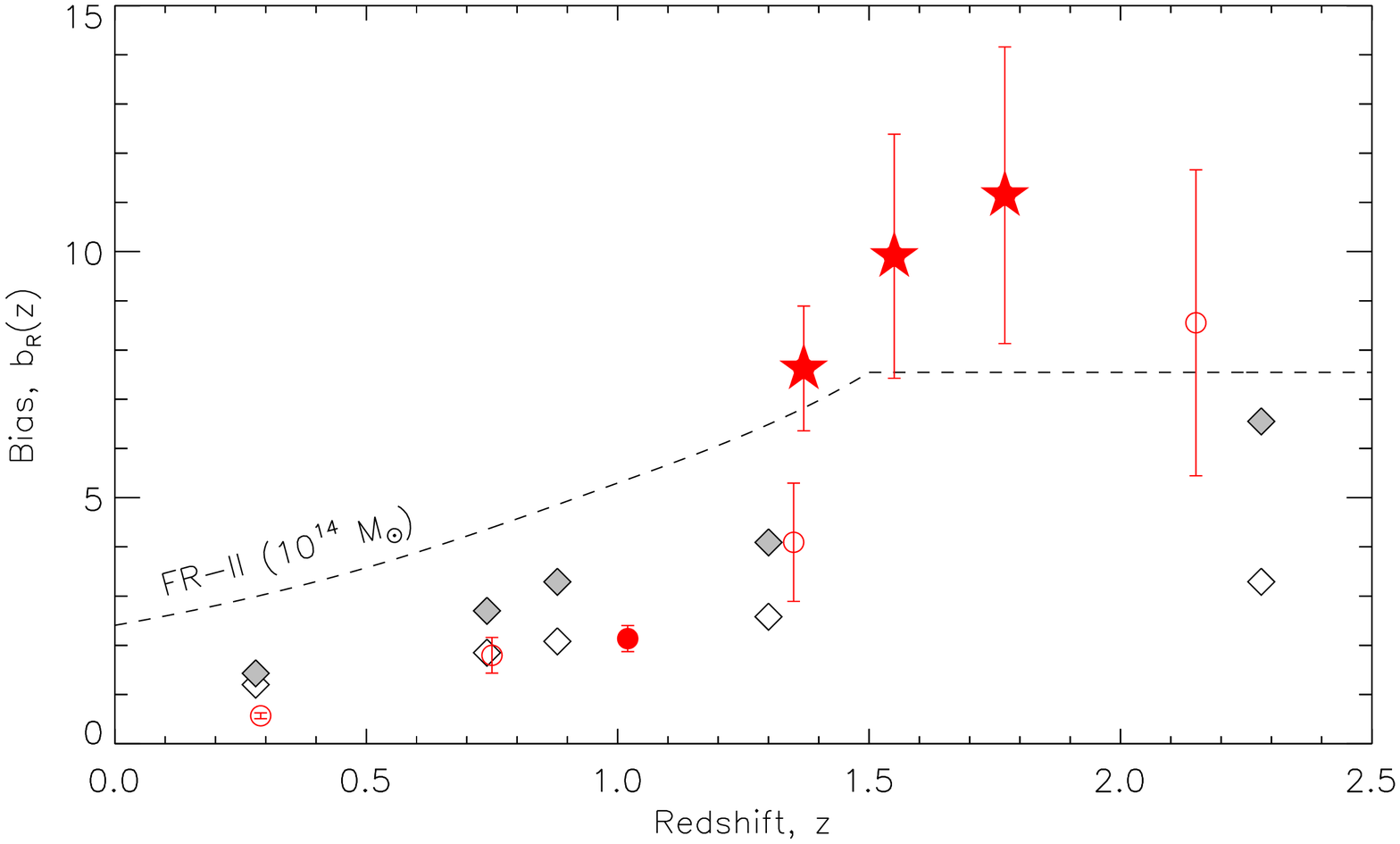}
   \caption{Linear bias of VIDEO-identified radio sources as a function of median redshift. Open circles correspond to the four independent redshift bins used while the filled circle is the bias for the full sample of 766 radio sources with $K_\textrm{s} < 23.5$ and $S_{1.4} > 90$ $\mu$Jy. Star symbols correspond to lower luminosity limits of $10^{23}$, $10^{23.5}$ and $10^{24}$ WHz$^{-1}$ from low to high redshift. The dashed line shows the FR-\textsc{II} bias adopted by \citet{wilman08} in the SKADS simulations, and the diamond symbols show the expected bias based on the SKADS prescriptions (\textit{open}) and with the FR-\textsc{I} halo mass increased to $10^{14}$ $M_\odot$ (\textit{filled}), matching the FR-\textsc{II}s. }
  	\label{biasR}
\end{figure*}

\subsection{The evolution of the bias of near-infrared selected galaxies}

In order to calculate the bias of the radio sources, we first calculate the angular autocorrelation function of $K_\textrm{s}$-selected VIDEO galaxies, finding the data to be fit very well with a single power law (Figure \ref{wIR}). This fit describes the data very well over almost two decades in angular scale, from $\theta=0.001$ to $0.08$ degrees. The correlation length and bias of these sources is then calculated using their redshift distribution (Figure~\ref{zbins}). We measure the evolution of the clustering of these sources in four bins of median redshift ranging from $z=0.3$ to $z=2.15$. These quantities are listed for the binned and full $z<4$ samples in Table \ref{IRpar} and shown in Figure \ref{biasIR}.

The power law slope of $w(\theta)$ is $\gamma-1 = 0.85$ for all but the highest redshift bin at $1.75<z<4$ where it rises to 0.98. While the angular clustering amplitude decreases with increasing redshift, Limber inversion gives a gently increasing correlation length, $r_0$ with increasing redshift, from $2.08 \pm 0.07$ $h^{-1}$Mpc at $z\sim0.33$ to $3.80^{+0.18}_{-0.19}$ $h^{-1}$Mpc at $z\sim2.16$. This corresponds to a bias increase from $0.59^{+0.02}_{-0.01}$ to $2.23\pm0.12$ and reflects the mix of galaxy populations present in a $K_\textrm{s}$-selected survey at these depths. The clustering of $K_\textrm{s}$-band selected galaxies as a function of galaxy mass and redshift will be studied in much greater detail in a subsequent paper. 

\subsection{The clustering and bias of faint radio sources}

In this work we are interested in the evolution in the bias of the faint radio source populations. We measure this using the cross-correlation function of the radio and $K_\textrm{s}$-selected galaxies.
Figure~\ref{wcross} shows the cross-correlation function of radio and $K_\textrm{s}$-selected galaxies for the full sample of galaxies. We also determine the cross-correlation in the four redshift bins as defined in the previous section (see Appendix). Furthermore, we impose lower limits on the radio luminosity of $10^{23}$, $10^{23.5}$ and $10^{24}$ WHz$^{-1}$, as each of these provides a slightly different sample of AGN with the fraction of normal star-forming galaxies gradually diminishing with radio luminosity (see Figure \ref{zbins}). Showing how this affects the measured clustering is important to gain fresh insight into the bias of AGN in an as yet uninvestigated luminosity and redshift range.  

The cross-correlation length and radio-infrared relative bias ($b_{K\textrm{R}}$), are calculated using the radio source redshift distribution and we then infer the radio bias using equation \ref{eq:radiobias2}. Figure \ref{biasR} shows the bias as a function of redshift for the radio sources. These quantities are also tabulated in Table~\ref{crosspar} for the various redshift bins and for the different luminosity limits.

We find a steeper power-law slope for the cross-correlation function than for the corresponding infrared autocorrelation function, with $\gamma - 1 > 1$ in all cases. Likewise, the radio-infrared correlation length is greater than the infrared galaxy correlation, giving a relative bias, $b_{K\textrm{R}}$, increasing even more strongly with redshift from $0.70^{+0.05}_{-0.05}$ to $3.60^{+0.99}_{-0.84}$ between the lowest and highest redshift bins at $z_\textrm{med} \sim 0.29$ and $2.15$ respectively. Accounting for the results of the infrared galaxy correlation function, this corresponds to a radio source bias of $0.57\pm0.06$ to $8.55\pm3.11$ over the same redshift range. 

Imposing a minimum radio luminosity criterion to our radio source sample, we find $r_0$ to be greater than that for the full sample, meaning that the higher-luminosity AGN are more strongly clustered than the general radio source population. The three different luminosity cuts result in similar correlation lengths ($r_0\sim5.5$ $h^{-1}$Mpc), with a slight decrease at the high luminosity end, but still within errors. The radio bias, however, increases as the radio luminosity increases, from $b_\textrm{R} = 7.62\pm1.27$ for $L>10^{23}$  WHz$^{-1}$(approximately double the bias for the full radio sample at $1<z<1.75$ with a similar median redshift) to $b_\textrm{R} = 11.14\pm3.01$ for $L>10^{24}$ WHz$^{-1}$. This increase is likely due purely to the increasing median redshift of higher luminosity sources, rather than a significantly more massive or more clustered sample.

\section[]{Discussion}\label{discussion}
	
\subsection{The clustering of near-infrared selected galaxies }

As the measurement of the clustering of the radio source population is dependent on our measurement of the clustering of the near-infrared galaxies through the cross-correlation function, it is informative to compare our results for the near-infrared galaxy clustering to other results in the literature.

Our results for the $K_\textrm{s}$-selected VIDEO sources, based on tightly constrained angular correlation functions (see Figures \ref{wIR} and \ref{zsamples}), show an increasing clustering strength with redshift, as would be expected for a population relaxing over cosmic time to follow the distribution of the underlying dark matter structure. However, comparison with similar studies of near-infrared galaxies suggests that we underestimate this clustering. 

\citet{furusawa11} investigated the mass-dependent clustering of a similarly derived sample of more than 50,000 $K<23.5$ galaxies from 0.63 deg$^2$ of the Subaru/XMM-\textit{Newton} Deep Survey and UKIRT Infrared Deep Sky Survey/Ultra Deep Survey (SXDS/UDS). While they do not show results for a combined sample across all redshifts, or indeed all masses, we find our redshift-binned $r_0$ to be lower than even their least clustered (lowest mass) galaxies at similar redshifts. 
Similarly, \citet{bielby13} find $r_0$ as a function of mass and redshift through the angular correlation function of a $z\lesssim 2$ galaxy sample from 2.4 deg$^2$ of the WIRCam Deep Survey (WIRDS) in $J$, $H$ and $K_\textrm{s}$ bands combined with optical data from the CFHTLS Deep fields. By deprojecting $w(\theta)$ analytically rather than using the Limber method, they find the correlation length to increase from $r_0 \sim 4.5$ $h^{-1}$Mpc at $z\sim0.5$ to $r_0 \sim 6.3$ $h^{-1}$Mpc at $z \sim 1.75$. 

Without reaching the same depth as ours and other authors' work, \citet{quadri07} use the Limber method to establish the correlation length and bias of a sample of $K<21$ (Vega magnitude) galaxies at $2<z<3.5$ using four $10^\prime \times 10^\prime$ fields of the MUSYC survey. Their bias at $z \sim 2.6$ of $3.3\pm0.5$ is slightly greater than we would expect extrapolating beyond our $z\sim2.16$ figure of $b=2.23\pm0.12$. We observe a slightly fainter population of sources, but Quadri et al. also show that there is little to no significant effect of limiting $K$-magnitude on clustering at these magnitudes. They do note the possible limitation, however, of fixing the slope parameter $\beta = \gamma-1 = 0.8$, as is done by \citet{furusawa11} (but not \citealt{bielby13}). The inset in Figure~\ref{wIR} shows that fixing the slope to 0.8 rather than fitting for it, we would artificially boost the amplitude parameter, $A$, and therefore $r_0$ and bias along with it. This partly addresses the difference between our highest-redshift result and that of \citet{quadri07} or even \citet{furusawa11}, but not \citet{bielby13}.

\citet{ichikawa07} provide severely limited constraints (due to a 24.4 arcmin$^2$ area) on the clustering of $K$-selected sources at fainter levels ($K<25$) using the MOIRCS Deep Survey in the GOODS-North region. Contrary to the shallower sample used by \citet{quadri07}, they find a significant decline in $r_0$ with $K$-band magnitude at $1<z<2$ and weaker evidence of such a decline at $2<z<4$. Even so, there is a roughly 1--2$\sigma$ disagreement in our results at $K<23.5$, more consistent with their results at $K<25$.

We have explored possible causes for discrepancy in our results compared with the aforementioned authors. Changing the range of our fits to $w(\theta)$ to match other work and fixing the slope to 0.8 (as is common practice where data are sparse), has little combined effect on the resulting correlation lengths and biases. The difference then, may lie in the redshift distributions used, which do impact the spatial clustering measures. In order to verify that our redshift distribution is robust, we have calculated $r_0$ for our data, but assuming the distribution of photometric redshifts from a similar sample taken from the deeper UltraVISTA survey \citep{mccracken12} over the COSMOS field, finding again that our results are not significantly altered. The redshift catalogue from \citet{muzzin13} includes photometry in 30 bands over the 1.62 deg$^2$ COSMOS/UltraVISTA field, giving redshifts with a catastrophic outlier fraction of just 1.6 per cent and whilst their distribution is not identical to that of the VIDEO sources, the consistent results suggest that we may be confident in our results with these redshifts. 


\subsection{Clustering of faint radio sources}

Our results show a strong evolution in the clustering and bias of faint radio sources with redshift. We also find that the more luminous subsamples are strongly biased, as would be expected if the dominant population were radio-loud AGN predominantly hosted by massive elliptical galaxies \citep[e.g.][]{jarvis01,dunlop03,herbert11}.

While \citet{wilman03} investigated the clustering of a radio sample down to 0.2 mJy, direct measurements of the clustering of radio sources below the mJy level are generally lacking in the present literature. However, some predictions for the bias of our sample may be taken indirectly from the SKA Design Study (SKADS; \citealt{wilman08,wilman10}) semi-empirical simulations of extragalactic radio continuum sources. These simulations provide a catalogue of radio sources over 400 deg$^2$ to a depth of 10 nJy in 15 radio and mid-infrared bands, as well as providing $K$-band magnitudes and classifications of 5 different radio source types: normal star-forming galaxies, starbursts, radio-quiet AGN, FR-\textsc{I} and FR-\textsc{II} type AGN. Each of these source types are attributed a fixed halo mass and a bias based on these masses and the \citet{mo96} bias model. 

By imposing radio luminosity cuts to our sample at $L_{1.4}>10^{23}$ WHz$^{-1}$ and higher, we effectively bias it towards AGN \citep{condon02}. The increasing minimum luminosity raises the median luminosity of these subsamples, and therefore the median redshift as the highest redshift sources must be extremely luminous to reach the radio detection limit. Within errors, and ignoring the imposed flattening of the SKADS bias prescription beyond $z=1.5$ \citep[see][ for more details]{wilman08}, these `AGN-only' results are consistent with the SKADS FR-\textsc{II} bias with a halo mass of $10^{14}$\,$M_\odot$. This suggests, in agreement with \citet{lindsay14}, that the typical halo mass of the more numerous FR-\textsc{I}s (we expect $\sim 6$ FR-\textsc{II}s in the 1~square degree that we use here) must be in this region as well. The increasing bias with luminosity (and therefore redshift), additionally, provides observational evidence that the truncation of the AGN bias models at $z=1.5$ in \citet{wilman08} is not justified. While the reason for this plateau is to prevent an unphysical rise in the model bias beyond redshifts where it can be constrained in the literature, we show that the bias continues to evolve to $z> 2$.

To investigate this, we use a SKADS sample catalogue of $S_{1.4} > 90\mu$Jy radio sources, giving a redshift distribution and proportions for each source population, comprising FR-\textsc{II} and FR-\textsc{I} radio AGN, radio-quiet quasars star-forming galaxies and starbursts. Weighting the five different bias models associated with these different populations by the relative proportions (counted directly from flags in the simulation output), we determine the expected bias for each of our data points shown in Figure~\ref{biasR}.
Furthermore, we show these predictions for the case where we assume the same (higher) halo mass for the FR-\textsc{I} sources as for the FR-\textsc{II} sources, as suggested by \citet{lindsay14}. This boosts the bias as expected, particularly at the higher redshifts, consistent with our upper two redshift bins, but also exacerbates an apparent underestimate at low redshift. 

Deviations from the SKADS predictions could be the result of miscalculating the proportions of AGN and normal galaxies -- a greater than expected fraction of AGN at high-redshift would appear to increase the observed bias -- but the luminosity functions used by \citet{wilman08} should be well-constrained so as to alleviate this potential issue. The halo masses of the different types of galaxy, and how they evolve with redshift however, are less well known. Assuming robust luminosity functions, our results would appear to suggest a less biased (i.e. less massive) population at low redshift and a more biased one at higher redshift, contrary to the single fixed mass model of SKADS.  Although we note that the cosmological volume sampled at low redshift is relatively small and as such we expect the clustering signal to be dominated by sample variance, which we do not consider here as we are most interested in the high-redshift measurements, whereas at low-redshift large-area surveys are more suitable.

\section[]{Conclusions}\label{conclusions}

We have made use of complementary optical and infrared data from CFHTLS and VIDEO to infer the spatial clustering of radio sources at the $>90$ $\mu$Jy level in a 1 square degree field (VIDEO-XMM3) through the angular cross-correlation function. With a $K_\textrm{s} < 23.5$ sample of $\sim$100,000 galaxies with reliable photometric redshifts out to $z\sim4$, we find the auto-correlation function, spatial correlation length, and linear bias of the full sample and four independent redshift bins. Furthermore, we use the cross-correlation of these sources with our deep radio sample of 766 galaxies to infer their bias. The results can be summarized as follows:

\begin{enumerate}
	\item The slope of $w(\theta)$ for infrared-selected galaxies is generally consistent ($\sim-0.85$) but slightly larger than the canonical value of $-0.8$ often assumed in the literature. This does not significantly affect our results, and we find an increasing bias and correlation length in qualitative agreement with similar infrared studies, but with significantly weaker clustering than found elsewhere due to a better constrained redshift distribution.
	\item By cross-correlating with this background of infrared galaxies, we are able to find a robust cross-clustering signal for the radio galaxies across 4 redshift bins up to a median redshift of $z=2.15$.
	\item Combining the auto- and cross-correlation function results, we are able to disentangle the clustering scales of the two galaxy populations to find a radio source bias which increases from $0.57\pm0.06$ at $z\sim0.29$ to $8.55\pm3.11$ at $z\sim2.15$.
	\item Placing lower limits on radio luminosity at $10^{23}$, $10^{23.5}$ and $10^{24}$ WHz$^{-1}$ effectively reduces the sample to AGN-dominated subsample, giving a high-redshift measurement of the AGN bias of $b(z=1.37)=7.62\pm1.27$, $b(z=1.55)=9.91\pm2.48$ and $b(z=1.77)=11.14\pm3.01$, respectively.
    \item The radio bias found at higher redshifts is greater than that expected by assuming the models used in the SKADS radio simulations. However, assigning a similar halo mass to the FR-\textsc{I} sources as assumed for the FR-\textsc{II} sources in our radio sample largely addresses this discrepancy. Indeed, results using the high-luminosity subsample appear to confirm that low-radio-luminosity AGN have a typical halo mass similar to that assigned to FR-\textsc{II} radio galaxies ($10^{14}$ $M_\odot$).
\end{enumerate}


\section*{Acknowledgements}
MJJ acknowledges the South Africa National Research Foundation Square Kilometre Array Project for financial support.
This work is based on observations made with ESO telescopes at the La Silla Paranal Observatory under programme ID: 179.A-2006 and on data products produced by the Cambridge Astronomy Survey Unit on behalf of the VIDEO consortium. It is also based on observations obtained with MegaPrime/MegaCam, a joint project of CFHT and CEA/DAPNIA, at the Canada–France–Hawaii Telescope (CFHT) which is operated by the National Research Council (NRC) of Canada, the Institut National des Science de l'Univers of the Centre National de la Recherche Scientifique (CNRS) of France and the University of Hawaii. This work is based in part on data products produced at TERAPIX and the Canadian Astronomy Data Centre as part of the Canada–France–Hawaii Telescope Legacy Survey, a collaborative project of NRC and CNRS.


\footnotesize{
\bibliographystyle{mn2e}
\bibliography{biblio}	
}


\appendix
\section[]{Additional Figures}

\begin{figure*}
  \centering
    \includegraphics[width=0.48\textwidth]{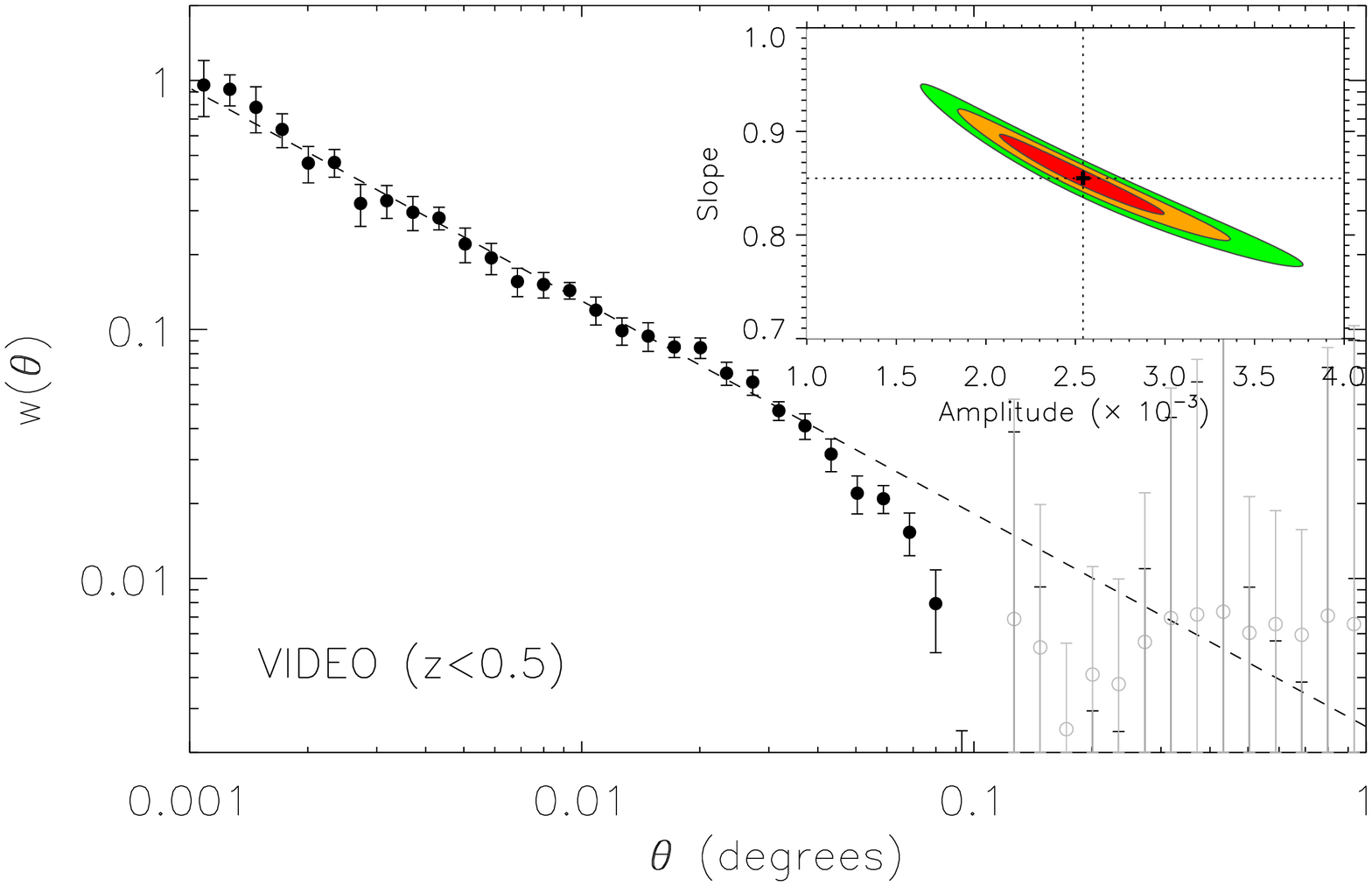} \qquad
    \includegraphics[width=0.48\textwidth]{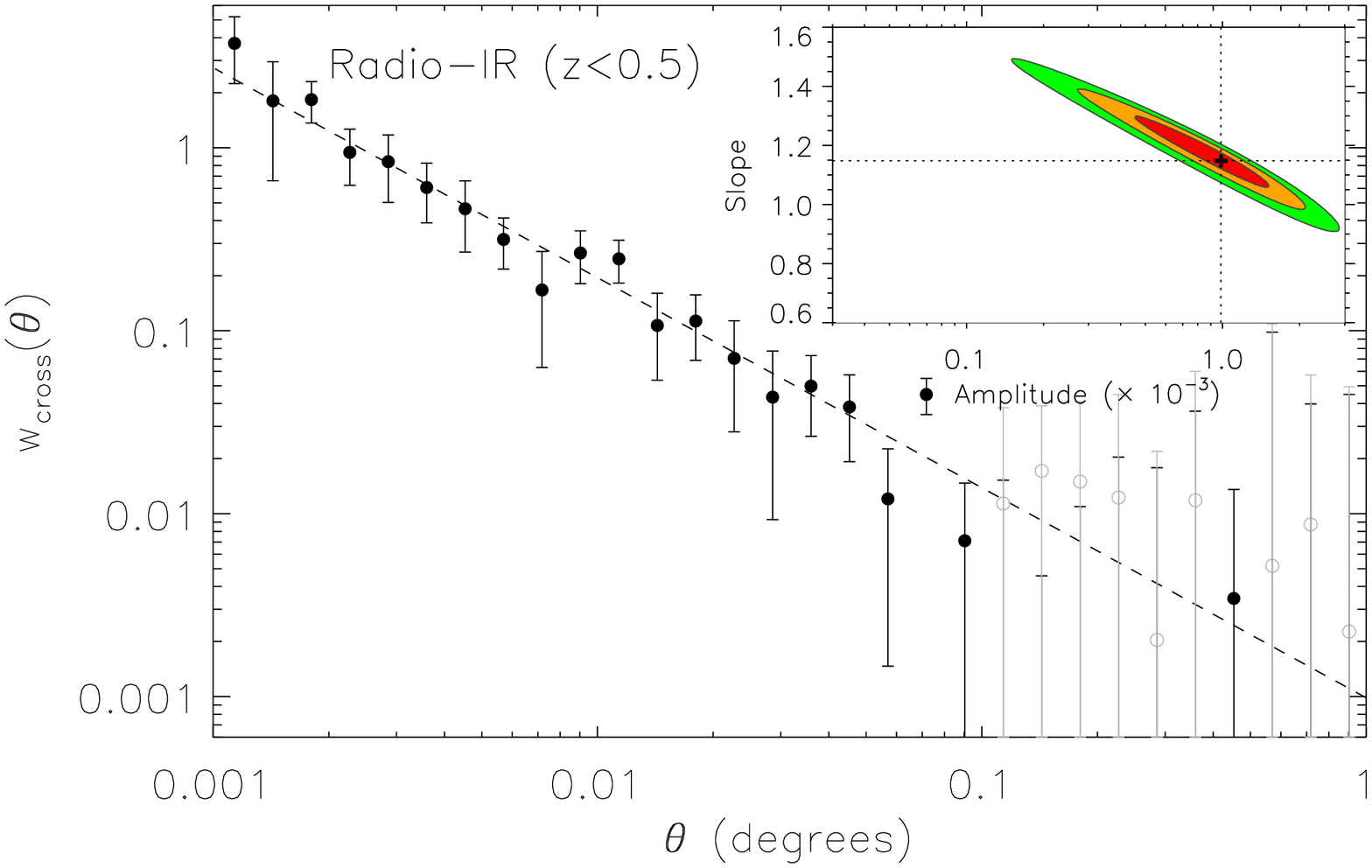} \\
        \includegraphics[width=0.48\textwidth]{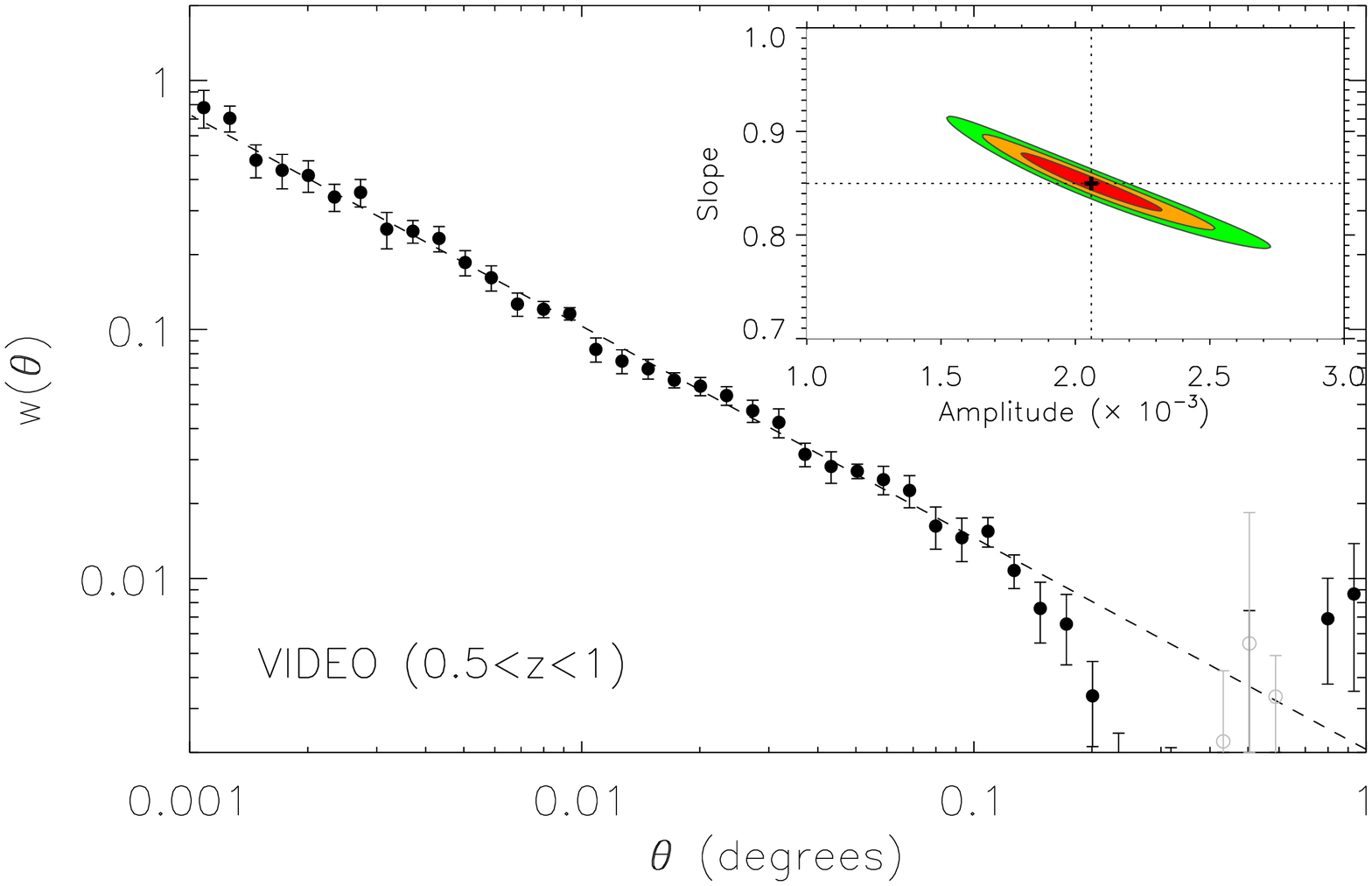} \qquad
    \includegraphics[width=0.48\textwidth]{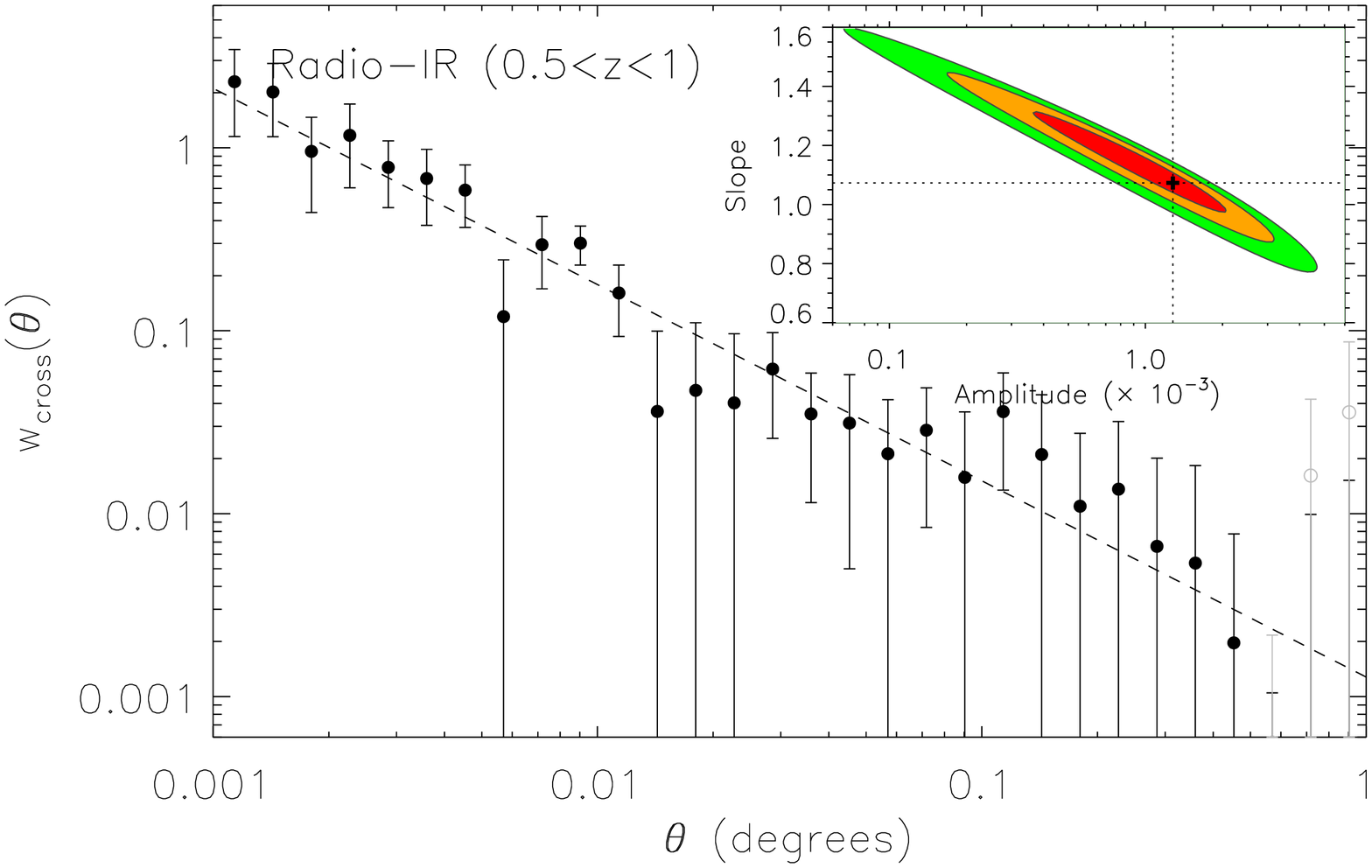} \\
        \includegraphics[width=0.48\textwidth]{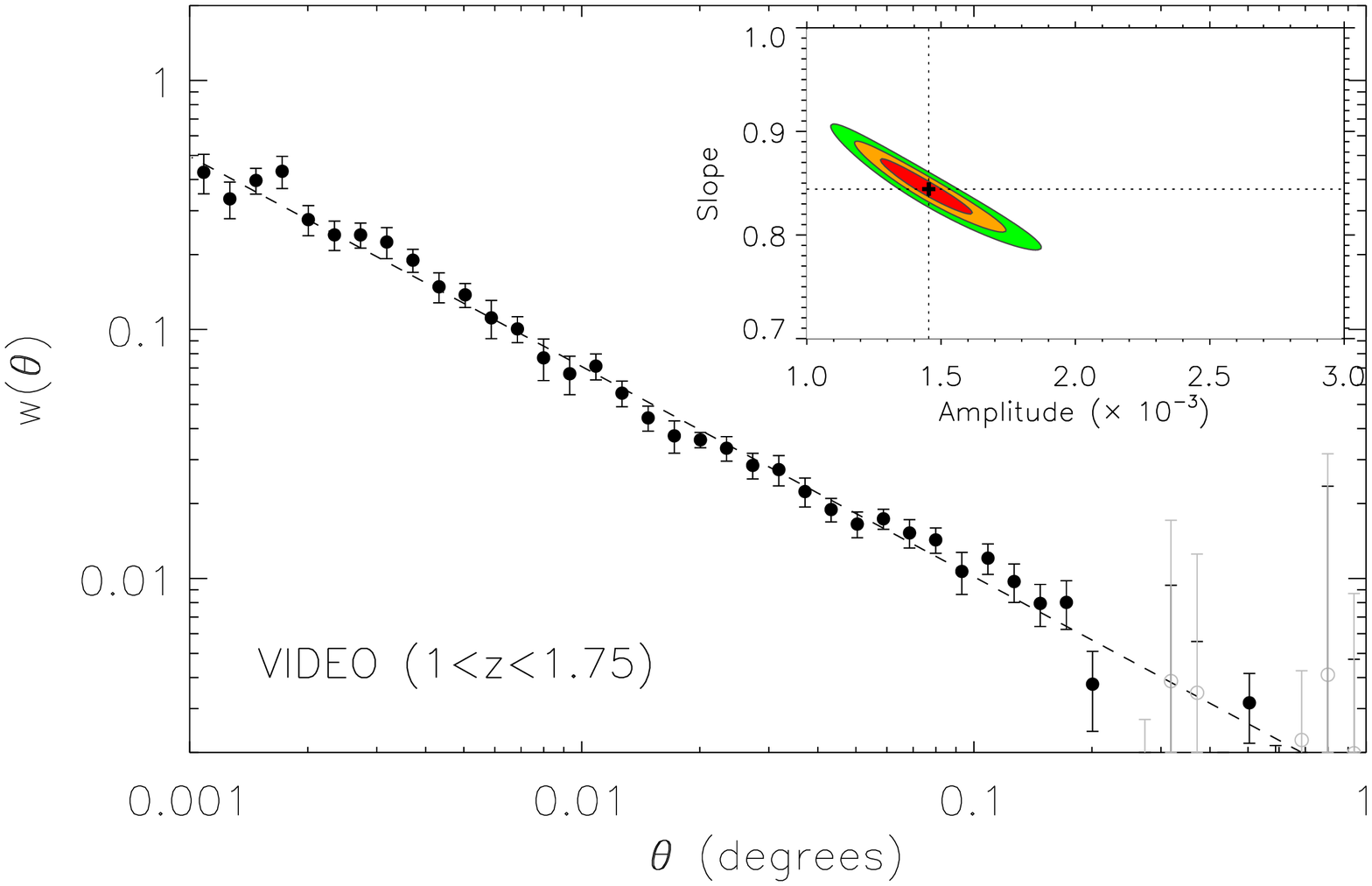} \qquad
    \includegraphics[width=0.48\textwidth]{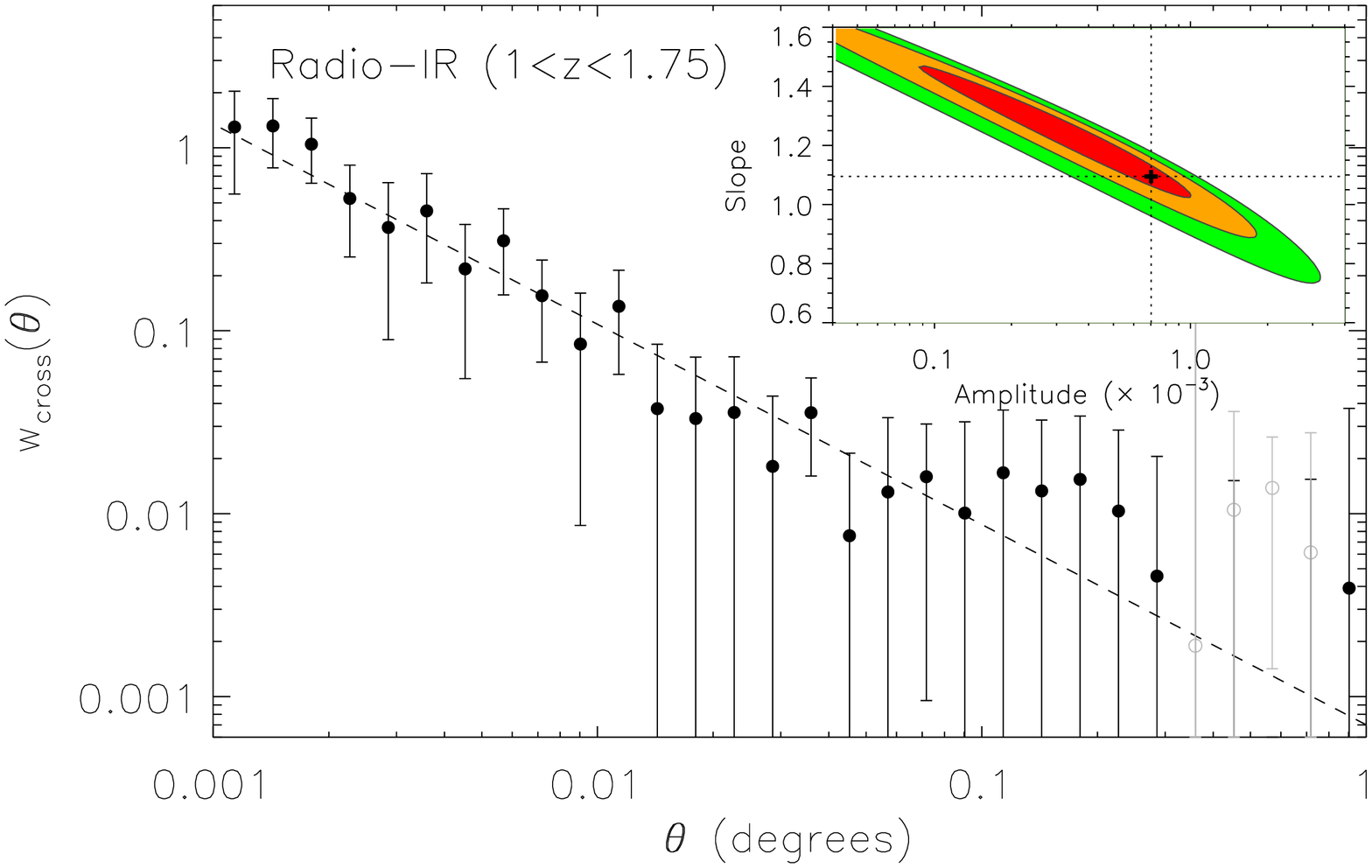} \\
        \includegraphics[width=0.48\textwidth]{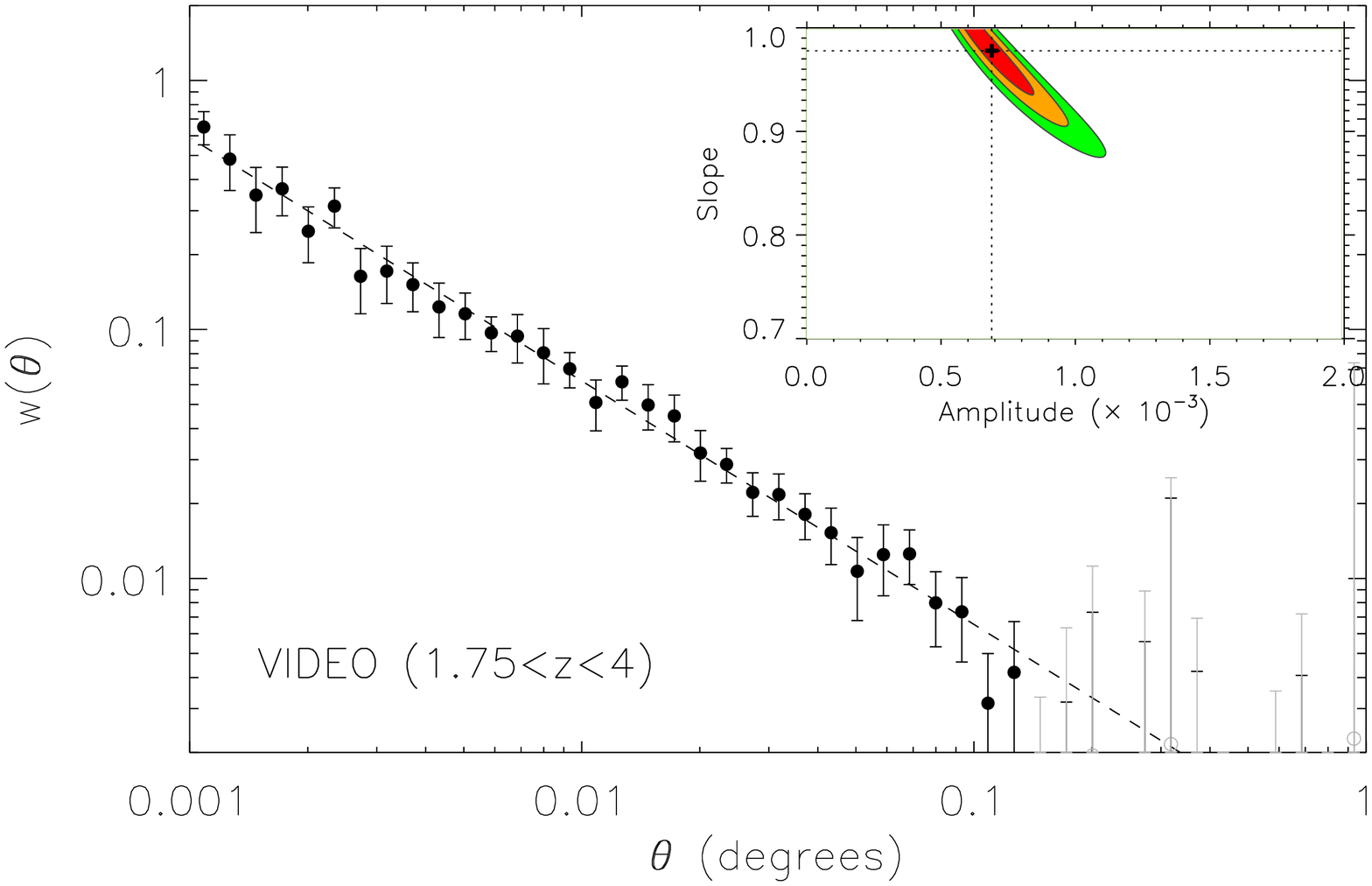} \qquad
    \includegraphics[width=0.48\textwidth]{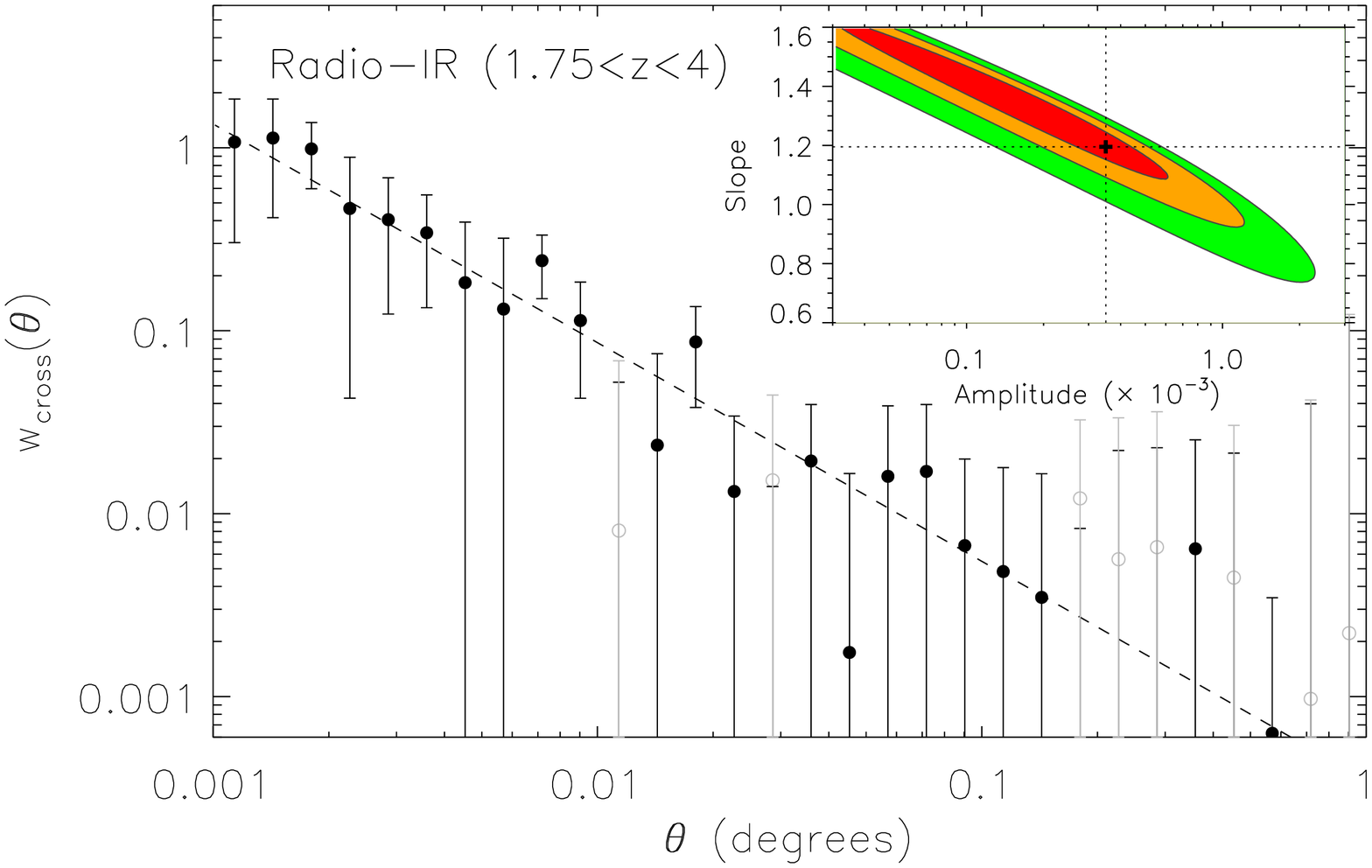} 
   \caption{The angular autocorrelation function (\textit{left}) of the $K_\textrm{s}<23.5$ VIDEO sources, and cross-correlation function (\textit{right}) with the radio counterparts (with bootstrap resampling errors) in 4 redshift bins. The dashed lines shows the best fit power-law and the inset plot shows $\chi^2$ parameter fits at 68, 90 and 95 per cent confidence levels. Points plotted in grey are the absolute values where $w_\textrm{cross} (\theta) < 0$.}
  \label{zsamples}
\end{figure*}

\begin{figure}
\centering
	\includegraphics[width=0.48\textwidth]{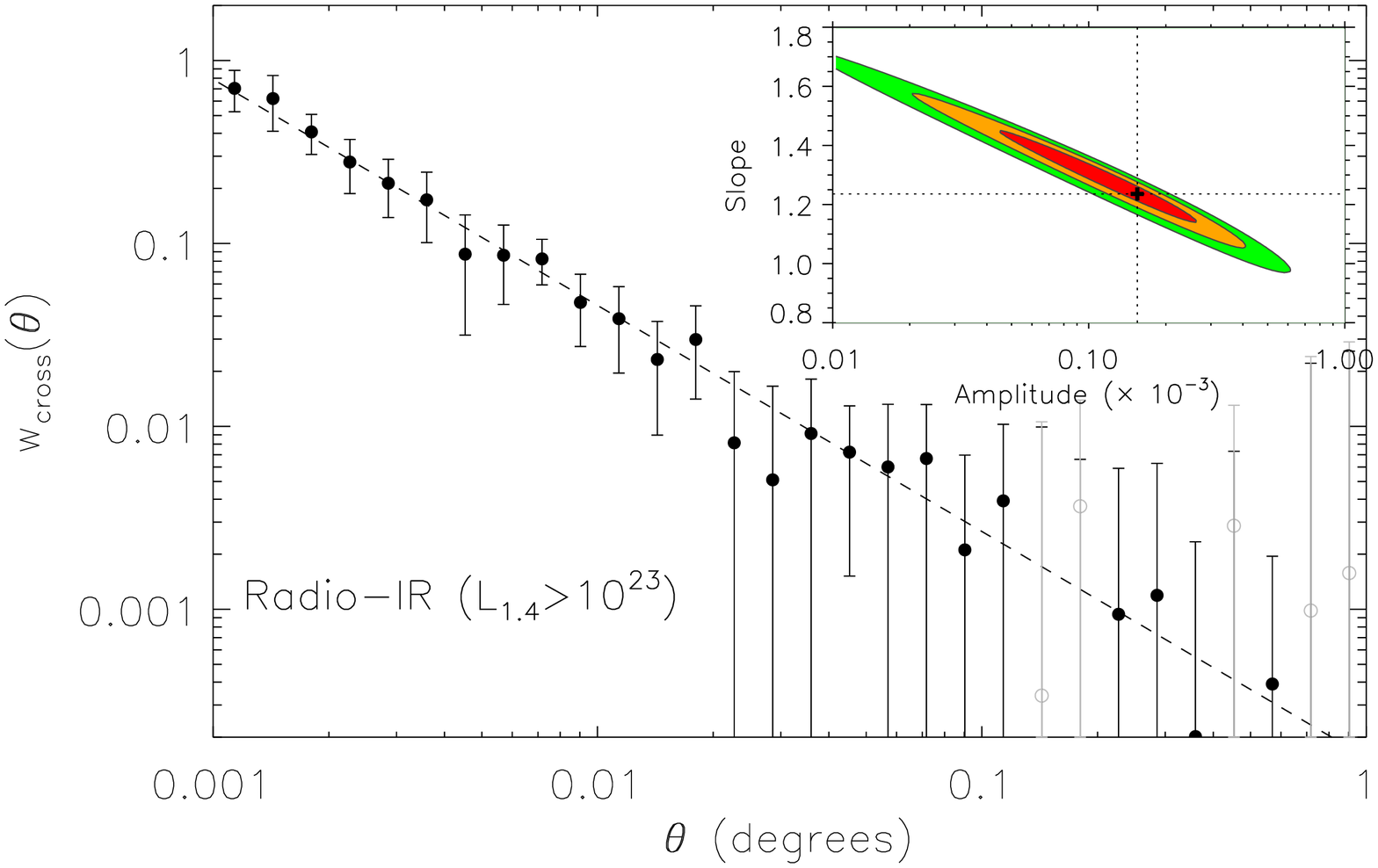} \\
	\includegraphics[width=0.48\textwidth]{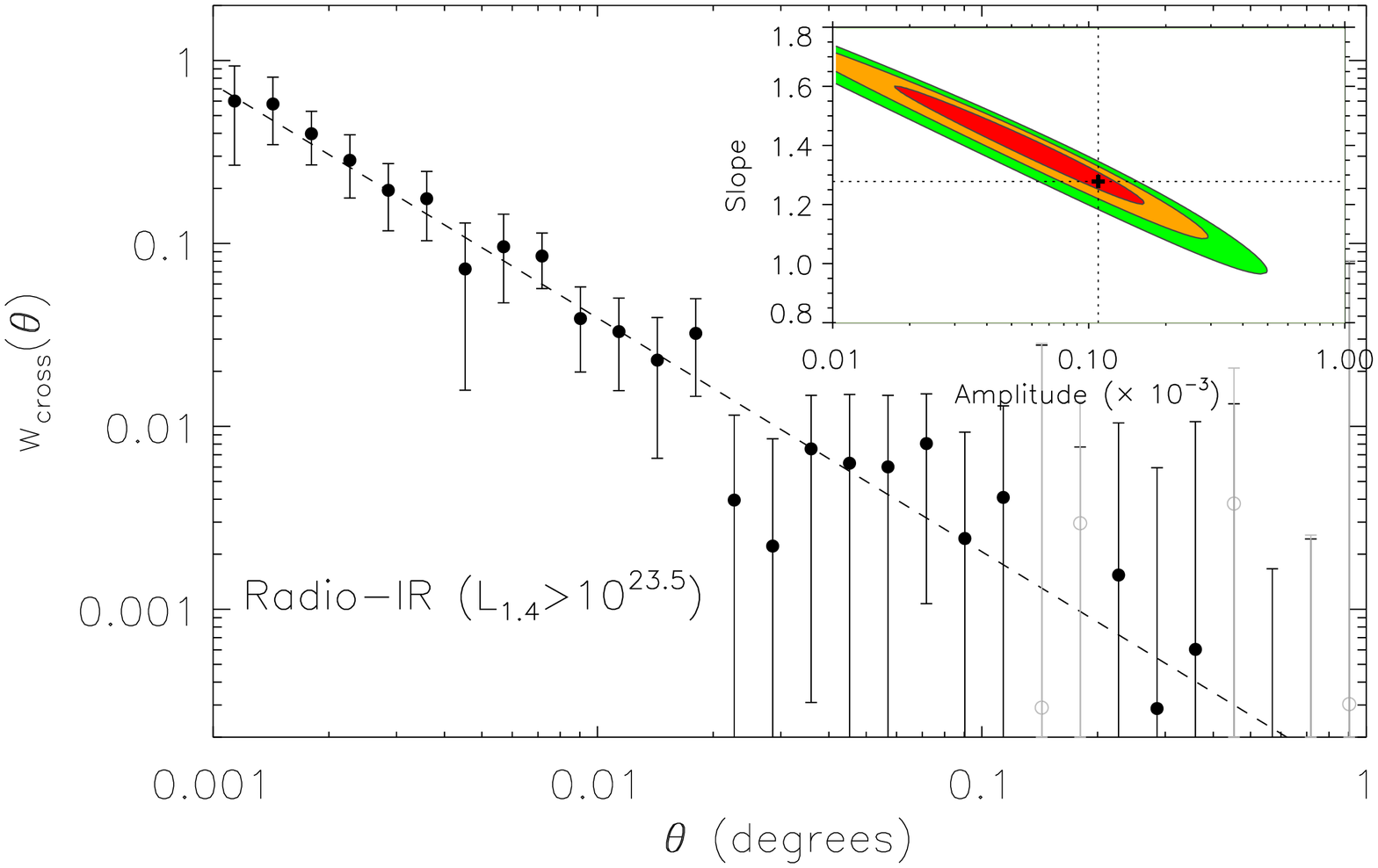} \\
	\includegraphics[width=0.48\textwidth]{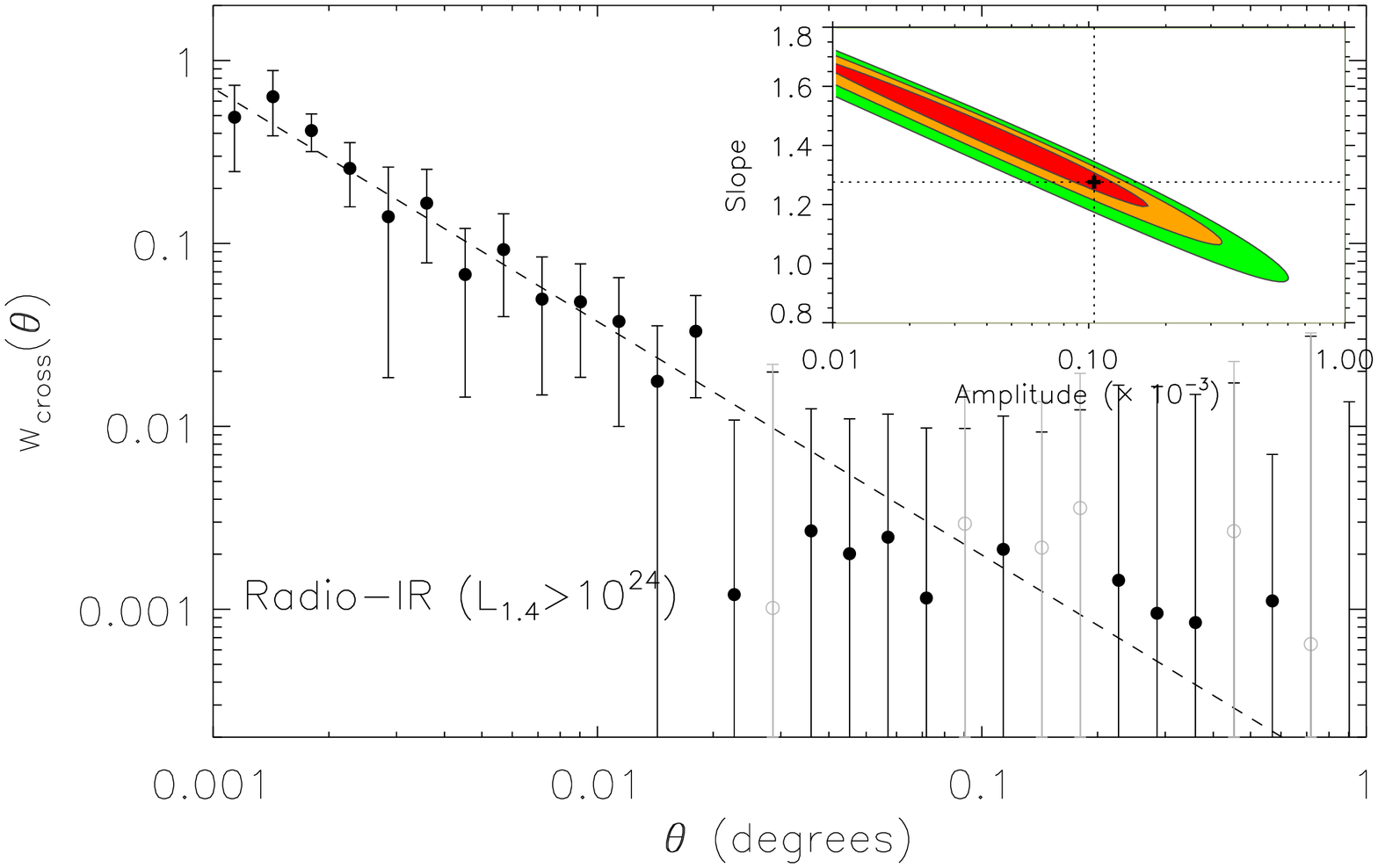} \\
\caption{The angular cross-correlation function of the $L_{1.4}>10^{23}$, $10^{23.5}$ and $10^{24}$ WHz$^{-1}$ radio counterparts with the VIDEO infrared sources (with bootstrap resampling errors). The dashed line shows the best fit power-law and the inset plot shows $\chi^2$ parameter fits at 68, 90 and 95 per cent confidence levels. Points plotted in grey are the absolute values where $w_\textrm{cross} (\theta) < 0$.}
\label{lsamples}
\end{figure}

\normalsize{
The auto- and cross-correlation functions for all of the redshift- and luminosity-limited sampled described in the text are shown here. Figure \ref{zsamples} shows the auto-correlation function of IR sources and cross-correlation function with radio sources, each in 4 separate redshift bins. Figure \ref{lsamples} shows the cross-correlation function of radio samples with different radio luminosity limits with the full $K_\textrm{s} < 23.5$ IR sample.
}

\label{lastpage}

\end{document}